\documentclass[12pt]{article}
\usepackage{amsmath}
\usepackage{amsfonts}
\usepackage{amssymb}
\usepackage{amsthm}
\usepackage{relsize}
\usepackage{dsfont}
\usepackage[latin1]{inputenc}
\usepackage[all]{xy}
\usepackage{graphicx}
\usepackage{latexsym}
\usepackage{color}
\usepackage{epsfig}
\usepackage{graphicx}
\usepackage{caption}
\usepackage{subcaption}
\usepackage{here}
\usepackage{float}
\usepackage{array}
\usepackage{tikz}
\usepackage{tikz-cd}
\usetikzlibrary{3d,arrows,calc,chains,matrix,positioning,scopes}

\def\l[{\left[}
\def\r]{\right]}

\newcommand{\tch}{{$\check{\mbox{C}}$}ech}

\def\ba{\begin{array}}
\def\ea{\end{array}}
\def\beq{\begin{equation}}
\def\eeq{\end{equation}}
\def\bea{\begin{eqnarray}}
\def\eea{\end{eqnarray}}

\newcommand{\coker}{\text{coker } \!}

\newcommand{\Hom}{\mathop{}\mathopen{}{\rm Hom}\!}

\newcommand{\egzz}{\, \stackrel{\mathbb{Z}}{{=}}\,}

\newtheorem{defi}{Definition}
\newtheorem{prop}{Property}

\begin{document}
\pagestyle{empty}
\setcounter{page}{0}
\hspace{-1cm}

\begin{center}
{\Large {\bf 3D Topological Models and Heegaard Splitting I: Partition Function}}%
\\ [1.5cm]

{\large F. Thuillier}

\end{center}

\vskip 0.7 truecm

{\it Laboratoire d'Annecy-le-Vieux de Physique Th\'eorique LAPTh, Universit\'e Savoie Mont Blanc, CNRS, BP 110, F-74941 Annecy, France}

\vspace{3cm}

\centerline{{\bf Abstract}}
The aim of this article is twofold: firstly, we show how to recover the smooth Deligne-Beilinson cohomology groups from a Heegaard splitting of a closed oriented smooth 3-manifold by extending the usual \tch-de Rham construction; secondly, thanks to the above and still relying on a Heegaard splitting, we explain how to compute the partition functions of the $U(1)$ Chern-Simons and BF theories.

\vspace{2cm}


\newpage
\pagestyle{plain} \renewcommand{\thefootnote}{\arabic{footnote}}

\section{Introduction}

For a smooth closed oriented $3$-manifold $M$, the Chern-Simons (CS) action of $M$ is usually given in the $SU(2)$ case under the form:
\bea
\label{SU2CS}
S_{CS} = k \frac{1}{8\pi^2} \int_M Tr\left( A \wedge d A + \tfrac{2}{3} A \wedge A \wedge A \right) \, ,
\eea
where $A$ is an $SU(2)$ gauge potential and $k$ a coupling constant. Under a gauge transformation, this action acquires a Wess-Zumino term which turns out to be, up to some $2\pi$ factor, an integer so that the quantum measure density $e^{i S}$ is gauge invariant quantity. This also requires $k$ to be a integer. It is now well known that the Partition Function of the corresponding Quantum Field Theory, written as a functional integral, coincides with the so-called Reshetikhin-Turaev (RT) invariant \cite{RT91}, and that the expectation values of Wilson loops define knot and links invariant of $M$, this invariant being closely related to Jones polynomial. In fact, no functional integration is performed when getting these results and its only heuristic arguments that yield the expressions of the RT invariant and Jones polynomial \cite{W89}. The same hold true for the so-called $SU(2)$ BF theory \cite{BBRT91} and its relation with the Turaev-Viro (TV) invariant \cite{TV92}.

Quite surprisingly, things are not so simple in the abelian, i.e. $U(1)$, case. The main reason is that, unlike $SU(2)$ principal bundles, $U(1)$ principal bundles over a smooth closed oriented $3$-manifold are not necessarily trivializable.  As a consequence, neither the CS action nor the BF action can be written in such a simple way as (\ref{SU2CS}). Deligne-Beilinson (DB) cohomology \cite{D71,B85} naturally appears in the abelian context because class of $U(1)$ gauge fields are nothing but DB cohomology classes, which eventually yields a proper definition for the CS and BF actions. In a long series of article, the determination of the partition function as well as of the expectation values of Wilson loops of the $U(1)$ CS \cite{GT2008,GT2013,GT2014} and BF \cite{MT1,MT2,MT3} topological quantum field theories on a closed $3$-manifold $M$ has been deeply investigated thanks to the use of (DB) cohomology. The partition functions were related with the abelian RT \cite{MOO92} and TV \cite{MT1} invariants respectively, expectation values were determined and related to abelian link invariants of $M$, and by using the Drinfeld center technics \cite{TVi13} surgery formulas were even obtained for the BF theory. In all this work, DB cohomology provides a way to identify the relevant finite number of degrees of freedom on which the various functional integrals have to be performed in order to recover the RT and TV invariants or the link invariants. The irrelevant infinite dimensional contribution is then eliminated thanks to an appropriate normalization. We refer to this as the ``standard approach".

In the non-abelian version of these theories, although the lagrangian are still DB cohomology classes, there is no such thing as a DB cohomology of non-abelian gauge fields that would allow us to identify the relevant finite dimensional contribution of the functional integral defining partition functions. Accordingly, there is no obvious normalization of the functional integrals defining these partition functions. Nevertheless, using a Heegaard splitting of $M$, the fact that the lagrangian is a DB class allows to rewrite classical CS invariants -- define as the CS action evaluated on flat connections -- as the sum of two surface terms, one typically representing colored intersections and the other being a Wess-Zumino term \cite{GMT17}. The colored intersections term is the only one surviving in the $U(1)$ case and gives rise to the correct expression of the classical $U(1)$ invariant.

In this article we investigate an extended use of a Heegaard splitting of a smooth closed oriented $3$-manifold $M$ in order to compute the partition functions of the $U(1)$ CS and BF theories. As we will see, the functional integral is performed on representatives of gauge classes rather than on classes themselves (which is the procedure in the standard approach). This approach is closer to the one used in Quantum Field Theory where a gauge fixing procedure has to be used. However, since the gauge fixing is performed  on a finite dimension sector only there is no need of a Faddeev-Popov term for this part of the functional integral. As to the infinite dimensional part it is once more eliminated via a normalization.

In a first section we will first recall some important facts concerning Heegaard splitting of oriented closed $3$-manifolds as well as basic data that we will need in the other sections. Then, we will explain how to recover the DB cohomology groups from a Heegaard splitting  point of view. Finally, we will introduce the notion of DB pairing and show how it is related to the linking form.

In a second section we will show how the CS and BF actions on $M$ can be written with the mean fields which are deeply related with the Heegaard splitting of $M$ and finally how the functional integral can be performed thus leading to the partition functions of the corresponding theories. We will check that the results coincide with those of the standard approach. Let us point out that the construction will be based on gauges fields of the handlebody on which the Heegaard splitting is based. This means that we could make the construction without referring to DB cohomology. However, it is DB cohomology which is behind the scene, this is why we decided to introduce it from the start thus taking advantage of this powerful mathematical structure.

As a final remark, let us recall that smooth DB cohomology is equivalent to Cheeger-Simons Differential Characters \cite{CS73} and Harvey-Lawson-Zweck Sparks \cite{HLZ03,HL01}. The results of the Sparks approach are intensively used in the present article.

All along this article, the symbol $\egzz$ will stand for equality modulo integers.

\section{Smooth DB cohomology}

Roughly speaking, a Heegaard splitting of a closed $3$-manifold $M$ is a closed surface $S$ which slits $M$ into two pretzels, i.e. handlebodies, $X_1$ and $X_2$ whose boundary is precisely $S$. It turns out that any closed and oriented $3$-manifold admits such a Heegaard splitting. By definition, the genus of $X_1$ and $X_2$ is the genus of their common boundary $S$, and the corresponding Heegaard splitting of $M$ is said to be of genus $g$.

Conversely, let $X$ be a genus $g$ handlebody, that is to say a compact $3$-dimensional manifold obtained by attaching $g$ handles to a ball $B^3$. The boundary of $X$ is a closed surface of genus $g$ that we denote by $\Sigma$. We provide $X$ with the standard ``outward" orientation which also induces an orientation on $\Sigma$. Let $\varphi:\Sigma \rightarrow \Sigma$ be a homeomorphism for which we set $\epsilon_\varphi := - \deg \varphi = \pm 1$. The gluing $X \cup_\varphi (\epsilon_\varphi X)$ of $X$ with $\epsilon_\varphi X$ according to $\varphi$, is an oriented closed $3$-manifold $M$ for which $X \cup_\varphi (\epsilon_\varphi X)$ is a Heegaard splitting, and any closed and oriented $3$-manifold is homeomorphic to such a Heegaard splitting.

Here, we consider the following alternative construction. We set $X_L = X$ and $X_R = - X$ and consider an orientation \textit{reversing} homeomorphism $f:\Sigma_L \rightarrow \Sigma_R$, where $\Sigma_L = \partial X_L = \Sigma$ and $\Sigma_R = \partial X_R = - \Sigma$. The Heegaard splittings of oriented closed $3$-manifolds we will consider are now $X_L \cup_f X_R$. As we will see, this way of doing will make signs easier to deal with.

The purpose of this section is to have all the necessary geometric ingredients before approaching the construction of the DB cohomology groups of $X_L \cup_f X_R$. While the content in this section might be well known to the reader, we think it is useful to write it down with some details. In the first two parts of these reminders, we recall important and useful results concerning $2$-dimensional closed surface and $3$-dimensional handlebodies, the former being the boundary of the latter. The third part is devoted to the description of chains, differential forms and de Rham currents of a Heegaard splitting $M = X_L \cup_f X_R$ from the point of view of the building data $X_L$, $X_R$ and $\Sigma_R$.

\subsection{Geometrical data}

Let $\Sigma$ be a closed surface of genus $g$ on which we choose a set $(\lambda_a^{\Sigma},\mu_b^{\Sigma})_{a,b=1, \cdots,g}$ of generators of its first homology group $H_1(\Sigma) \simeq \mathbb{Z}^{2g} $. The $1$-cycles $\lambda_a^{\Sigma}$ are referred to as \textbf{longitudes} and the $1$-cycles $\mu_b^{\Sigma}$ as \textbf{meridians}. As in the Heegaard construction the surface $\Sigma$ is the boundary of an oriented handlebody $X$, this surface inherit an orientation from the one of $X$. With respect to this orientation, the longitudes and meridians are chosen in such a way that they fulfilled:
\begin{equation}
\label{intersection}
\lambda_a^{\Sigma} \odot_\Sigma \mu_b^{\Sigma} := (\lambda_a^{\Sigma} \pitchfork \mu_b^{\Sigma})^{\sharp_0} = \delta_{ab} \, ,
\end{equation}
and zero in all other cases. The pairing $\odot_\Sigma$ is usually referred to as the \textbf{intersection form} on $\Sigma$. It is obtained by combining the transverse intersection $\pitchfork$ with the degree operator $\sharp_0:C_0(\Sigma) \rightarrow \mathbb{Z}$ which is an extension of the boundary operator to $0$-chains so that $\sharp_0 \circ \partial = 0$. Typically, $(\pm x)^{\sharp_0} := \pm 1$ for any point $x$ seen as a positively oriented $0$-chain of $\Sigma$. On a path connected manifold a $0$-chain is the boundary of a $1$-chain if and only if it has degree zero.
The set of $p$-chains, $p$-cycles and $p$-boundaries of $\Sigma$ are respectively denoted by $C_p(\Sigma)$, $Z_p(\Sigma)$ and $B_p(\Sigma)$, with $\partial_\Sigma$ the boundary operator, and $(C_\bullet(\Sigma),\partial_\Sigma)$ the corresponding chain complex of $\Sigma$.

As $\pitchfork$ is skew-symmetric on $1$-cycles of $\Sigma$, the product $\odot$ is so too, so that we have:
\bea
\label{skewsymodot}
\mu_b^{\Sigma} \odot_\Sigma \lambda_a^{\Sigma} = - \lambda_a^{\Sigma} \odot_\Sigma \mu_b^{\Sigma} = - \delta_{ab} \, .
\eea
Let us recall that $H_0(\Sigma) \simeq \mathbb{Z}$ since $\Sigma$ is connected, and that $H_2(\Sigma) \simeq \mathbb{Z}$. We can see $\Sigma$ itself as a generator of $H_2(\Sigma)$.

In order to obtain the cohomology groups $H^p(\Sigma)$ of $\Sigma$ we can just apply Poincar\'e duality theorem which yields the following set of isomorphisms:
\bea
\label{PLdualitySigma}
\left\{ \begin{aligned}
& H^0(\Sigma) \simeq H_2(\Sigma) \simeq \mathbb{Z} \\
& H^1(\Sigma) \simeq H_1(\Sigma) \simeq \mathbb{Z}^{2g} \\
& H^2(\Sigma) \simeq H_0(\Sigma) = \mathbb{Z}
\end{aligned} \right. \, .
\eea

The vector space of smooth $p$-forms on $\Sigma$ is denoted by $\Omega^p(\Sigma)$ and the corresponding de Rham complex by $\left(\Omega^\bullet(\Sigma), d_\Sigma \right)$. The subspace of closed $p$-forms of $\Sigma$ is then denoted by $\Omega^p_0(\Sigma)$ and the one of closed $p$-forms with integral periods by $\Omega^p_{\mathbb{Z}}(\Sigma)$. To any $1$-cycle $z$ of $\Sigma$ we can associate $j_z^\infty \in \Omega^1_{\mathbb{Z}}(\Sigma)$ which has compact support in a tubular neighborhood of $z$ as close to $z$ as we want and such that:
\bea
\label{poincaredualrep}
\forall \alpha \in \Omega^p_0(\Sigma) \; , \; \; \; \; \oint_{z} \alpha = \oint_{\Sigma} \alpha \wedge j_z^\infty \, .
\eea
Such a $1$-form is usually referred to as a representative of the Poincar\'e dual of the homology class of $z$ \cite{BT82}. The above relation actually relates de Rham cohomology with singular homology of $\Sigma$, in particular $j_z^\infty$ is note unique.

Let $j_{\lambda^\Sigma_a}^\infty$ and $j_{\mu^\Sigma_a}^\infty$ be smooth representative of $\lambda_a^{\Sigma}$ and $\mu_b^{\Sigma}$. It is always possible to chose these representative in such a way that:
\bea
\label{smoothintersection}
\delta_{ab} = \int_\Sigma j_{\lambda_a^{\Sigma}}^\infty \wedge j_{\mu_b^{\Sigma}}^\infty = \oint_{\mu_b^{\Sigma}} j_{\lambda_a^{\Sigma}}^\infty  = -  \oint_{\lambda_b^{\Sigma}} j_{\mu_a^{\Sigma}}^\infty = - \int_\Sigma j_{\mu_b^{\Sigma}}^\infty \wedge j_{\lambda_a^{\Sigma}}^\infty \, ,
\eea
thus recovering the intersection form of $\Sigma$.

From the perspective of the Heegaard construction we are going to consider, let $\Sigma_L$ and $\Sigma_R$ be two copies of $\Sigma$ with opposite orientation\footnote{In the Heegaard construction, the orientation of each is inherited from the one of the handlebody of which each copy is the boundary.}, and let $f:\Sigma_L \rightarrow \Sigma_R$ be an orientation reversing diffeomorphism. We need to specify what boundary the longitudes and meridians are considered on. Once done, we still assume that:
\bea
\label{conventionSigmaR}
\lambda_a^{\Sigma_L} \odot_{\Sigma_L} \mu_b^{\Sigma_L} = \delta_{ab} = \lambda_a^{\Sigma_R} \odot_{\Sigma_R} \mu_b^{\Sigma_R} \, .
\eea

The diffeomorphism $f:\Sigma_L \rightarrow \Sigma_R$ induces a morphism $f:Z_p(\Sigma_L) \rightarrow Z_p(\Sigma_R)$, still denoted by $f$ as there is no real risk of confusion. This morphism on its turn gives rise to an isomorphism $\hat{f}:H_1(\Sigma_L) \rightarrow H_1(\Sigma_R)$ which is completely determined by giving the  action of $f$ on $\lambda_a^{\Sigma_L}$ and $\mu_b^{\Sigma_L}$. This action takes the generic form:
\begin{equation}
\label{images}
\left\{ \begin{array}{l}
f(\lambda_a^{\Sigma_L}) = \sum\limits_{b=1}^g {{r_{ab}}} \, \lambda_b^{\Sigma_R} + \sum\limits_{b=1}^g {{s_{ab}}} \, \mu _b^{\Sigma_R} + \partial \varphi_a^{\Sigma_R} \\
f(\mu _a^{\Sigma_L}) = \sum\limits_{b=1}^g {{p_{ab}}} \, \lambda_b^{\Sigma_R} + \sum\limits_{b=1}^g {{q_{ab}}} \, \mu_b^{\Sigma_R} + \partial \psi_a^{\Sigma_R}
\end{array} \right. \, ,
\end{equation}
where $\varphi_a^{\Sigma_R}$ and $\psi_a^{\Sigma_R}$ are 2-chains of $\Sigma_R$. The transposed matrices of the matrices $(r_{ab})$, $(s_{ab})$, $(p_{ab})$ and $(q_{ab})$, respectively denoted by $P$, $Q$, $R$ and $S$, are matrices that each represent a morphism of $\mathbb{Z}^g$ into itself. The matrix representing the automorphism of $\mathbb{Z}^{2g}$ induced by $\hat{f}:H_1(\Sigma_L) \rightarrow H_1(\Sigma_R)$ takes form:
\begin{equation}
\label{matrixMf}
\mathcal{M}_{\hat{f}} =
\left( {\begin{array}{*{20}{c}}
R&P\\
S&Q
\end{array}} \right) \, .
\end{equation}
This matrix is such that $det \mathcal{M}_{\hat{f}} = \deg f = -1$, which is inferred from the consistency condition:
\bea
\label{consistencySigma}
f(\lambda_a^{\Sigma_L}) \odot_{\Sigma_R} f(\mu_b^{\Sigma_L}) = (\deg f) \, (\lambda_a^{\Sigma_L} \odot_{\Sigma_L} \mu_b^{\Sigma_L}) \, .
\eea

The diffeomorphism $f$ induces an isomorphism $f_*:\Omega^p(\Sigma_L) \rightarrow \Omega^p(\Sigma_R)$, usually referred to as the push-forward mapping. Then, the relations (\ref{images}) can be expressed in terms of $j_{\lambda^{\Sigma_L}_a}^\infty$ and $j_{\mu^{\Sigma_L}_a}^\infty$, $j_{\lambda^{\Sigma_R}_a}^\infty$ and $j_{\mu^{\Sigma_R}_a}^\infty$ by simply replacing meridians and longitudes by their smooth representatives. In doing so, the contributions $\partial \varphi_a^{\Sigma_R}$ and $\partial \psi_a^{\Sigma_R}$ must be replaced by smooth exact contributions $d{\kern 1pt} j_{\varphi_a^{\Sigma_R}}^\infty$ and $d{\kern 1pt} j_{\psi_a^{\Sigma_R}}^\infty$. As to the consistency condition (\ref{consistencySigma}), it takes the form:
\bea
\label{consistencySigma2}
\oint_{\Sigma _R} f_* j_{\lambda_a^{\Sigma_L}}^\infty \wedge f_* j_{\mu_b^{\Sigma_L}}^\infty  = (\deg f) \oint_{\Sigma _L} j_{\lambda_a^{\Sigma_L}}^\infty \wedge j_{\mu_b^{\Sigma_L}}^\infty \, .
\eea

The submatrix $P$ will be of particular interest. We already know that $P = (p)^\dag$. The matrix elements of $(p)$ can be obtained thanks to the intersection product since:
\bea
p_{ab} = f(\mu_a^{\Sigma_L}) \odot_{\Sigma_R} \mu_b^{\Sigma_L} \, .
\eea
Let us introduce the inverse of the relations (\ref{images}):
\begin{equation}
\label{inversimages}
\left\{ \begin{array}{l}
f^{-1}(\lambda_a^{\Sigma_R}) = \sum\limits_{b=1}^g {{\tilde{r}_{ab}}} \, \lambda_b^{\Sigma_L} + \sum\limits_{b=1}^g {{\tilde{s}_{ab}}} \, \mu _b^{\Sigma_L} + \partial \tilde{\varphi}_a^{\Sigma_L} \\
f^{-1}(\mu _a^{\Sigma_R}) = \sum\limits_{b=1}^g {{\tilde{p}_{ab}}} \, \lambda_b^{\Sigma_L} + \sum\limits_{b=1}^g {{\tilde{q}_{ab}}} \, \mu_b^{\Sigma_L} + \partial \tilde{\psi}_a^{\Sigma_L}
\end{array} \right. \, ,
\end{equation}
Then, we have:
\bea
\begin{aligned}
p_{ba} = & \; f(\mu_b^{\Sigma_L}) \odot_{\Sigma_R} \mu_a^{\Sigma_L} = (\deg f) \left( \mu_b^{\Sigma_L} \odot_{\Sigma_L} f^{-1}(\mu_a^{\Sigma_L}) \right) \\
= & \; - \mu_b^{\Sigma_L} \odot_{\Sigma_L} \left( \sum_{c=1}^{g} \tilde{p}_{ac} \lambda_c^{\Sigma_L} + \sum_{c=1}^{g} \tilde{q}_{ac} \mu_c^{\Sigma_L} \right) \\
= & \; \tilde{p}_{ab}
\end{aligned} \, ,
\eea
which means that:
\bea
(\tilde{p}) = (p)^\dag \, .
\eea
By repeating this argument with the products $\lambda_a^{\Sigma_R} \odot f(\mu_b^{\Sigma_L})$, $\mu_a^{\Sigma_R} \odot f(\lambda_b^{\Sigma_L})$ and $\lambda_a^{\Sigma_R} \odot f(\lambda_b^{\Sigma_L})$ we obtain the relations:
\bea
(\tilde{q}) = - (r)^\dag \; \; \; , \; \; \; (\tilde{r}) = - (q)^\dag \; \; \; , \; \; \; (\tilde{s}) = (s)^\dag \, .
\eea
If we introduce the matrices $\tilde{P}$, $\tilde{Q}$, $\tilde{R}$ and $\tilde{S}$ such that:
\bea
\mathcal{M}_{\hat{f}^{-1}} =
\left( {\begin{array}{*{20}{c}}
\tilde{R} & \tilde{P} \\
\tilde{S} & \tilde{Q}
\end{array}} \right) \, ,
\eea
then all these computations show that:
\bea
\label{inversmatrixMf}
\mathcal{M}_{\hat{f}^{-1}} =
\left( {\begin{array}{*{20}{c}}
-Q^\dag & P^\dag \\
S^\dag & -R^\dag
\end{array}} \right) =
\mathcal{M}_{\hat{f}}^{-1} \, .
\eea
From the relation $\mathcal{M}_{\hat{f}^{-1}} \mathcal{M}_{\hat{f}} = \mathds{1} = \mathcal{M}_{\hat{f}} \mathcal{M}_{\hat{f}^{-1}}$ we deduce a full set of very usefull relations:
\bea
\label{mainpropPQRS}
\begin{aligned}
Q^\dag P = P^\dag Q \; \; \; , \; \; \; P^\dag S - Q^\dag R = 1  \; \; \; , \; \; \; S^\dag R = R^\dag S \, , \\
R P^\dag = P R^\dag \; \; \; , \; \; \; S P^\dag - Q R^\dag = 1  \; \; \; , \; \; \; S Q^\dag = Q S^\dag \, ,
\end{aligned}
\eea
some of which will play an important role in the sequel.

\vspace{0.5cm}

Let us consider a $3$D handlebody $X$ with boundary $\partial X$ such that $i_\Sigma: \Sigma \rightarrow \partial X$ the canonical embedding. Each meridian $\mu_a^\Sigma$ can be sent into a meridian $i_\Sigma(\mu_a^\Sigma)$ of $\partial X$ which is actually a boundary in $X$. In order to lighten notations and since there is no real risk of confusion, we denote $\mu_a^\Sigma$ the meridians on $\partial X$. Let us consider a set $\left\{D_a\right\}_{a=1,\cdots,g}$ of disks in $X$ such that:
\bea
\label{meridiandisks}
\partial D_a = \mu_a^\Sigma .
\eea
These disks will be referred to as \textbf{meridian disks} of $X$. Similarly, each longitude $\lambda_a^\Sigma$ defines a longitude of $\partial X$ that we will also write $\lambda_a^\Sigma$. None of these longitudes is a boundary. However, let us homotopically move each $\lambda_a^\Sigma$ into $X$ so as to generate the so called \textbf{core cycles} of $X$. These $1$-cycles will be denoted $\lambda_a$. Then, we consider a set $\left\{A_a\right\}_{a=1,\cdots,g}$ of annuli of $X$ such that:
\bea
\label{meridiandisks}
\partial A_a = \lambda_a - \lambda_a^\Sigma .
\eea
Equations (\ref{meridiandisks}) and (\ref{meridiandisks}) reflect the fact that the core $1$-cycles $\lambda_a$ are generators of $H_1(X) \simeq \mathbb{Z}^g$ whereas the meridian disks $D_a$ are generator of $H_2(X,\partial X) \simeq \mathbb{Z}^g$. Theses equations also reflect the fact that $H_2(X) = 0$ and $H_1(X,\partial X) = 0$. It is quite obvious that $X$ is a boundary in $X$, which implies that $H_3(X) = 0$ whereas it is a generator of $H_3(X,\partial X) \simeq \mathbb{Z}$. Finally, any point of $X$ is a generator of $H_0(X) \simeq \mathbb{Z}$ but is a boundary in $(X,\partial X)$ which reflect the fact $H_0(X,\partial X) =0$.

The Poincar\'e-Lefschetz duality theorem \cite{D72} yields the following set of isomorphisms between homology and cohomology groups of $X$ and $(X,\partial X)$:
\bea
\label{PLDisomorphisms}
\left\{ \begin{gathered}
  {H^0}\left( X \right) \simeq {H_3}\left( {X,\partial X} \right) \simeq \mathbb{Z} \hfill \\
  {H^1}\left( X \right) \simeq {H_2}\left( {X,\partial X} \right) \simeq {\mathbb{Z}^g} \hfill \\
  {H^2}\left( X \right) \simeq {H_1}\left( {X,\partial X} \right) = 0 \hfill \\
  {H^3}\left( X \right) \simeq {H_0}\left( {X,\partial X} \right) = 0 \hfill \\
\end{gathered}  \right.\quad \quad ,\quad \quad \left\{ \begin{gathered}
  {H^0}\left( {X,\partial X} \right) \simeq {H_3}\left( X \right) = 0 \hfill \\
  {H^1}\left( {X,\partial X} \right) \simeq {H_2}\left( X \right) = 0 \hfill \\
  {H^2}\left( {X,\partial X} \right) \simeq {H_1}\left( X \right) \simeq {\mathbb{Z}^g} \hfill \\
  {H^3}\left( {X,\partial X} \right) \simeq {H_0}\left( X \right) = 0 \hfill \\
\end{gathered}  \right. \, .
\eea

The meridian disks $D_a$ and the core $1$-cycles $\lambda_a$ fulfill the following intersection property:
\bea
\label{Dandlambdaconstraint}
\lambda_a \odot_X D_b =  - \delta_{ab} = - \lambda^\Sigma_a \odot_\Sigma \mu^\Sigma_b \, ,
\eea
which extends to $X$ the intersection property (\ref{conventionSigmaR}). In this case the intersection form $\odot_X$ is commutative (because $\pitchfork$ is), and we have:
\bea
\label{odotofDlambda}
D_b \odot_X \lambda_a = \lambda_a \odot_X D_b \, .
\eea

Last but not least, we can associate \cite{dR55} to each core $1$-cycle $\lambda_a$, to each meridian disk $D_a$ and to each longitude annulus $A_a$, some smooth forms $j_{\lambda_a}^\infty$, $j_{D_a}^\infty$ and $j_{A_a}^\infty$ such that:
\begin{enumerate}
\label{smoothrepproperty}
\item $j_{\lambda_a}^\infty$ is a smooth representative of $\lambda_a$,
\item $d{\kern 0.3pt} j_{D_a}^\infty = 0$ and $d{\kern 0.3pt} j_{A_a}^\infty = j_{\lambda_a}^\infty$,
\item the restriction to $\partial X$ of $j_{D_a}^\infty$ and $j_{A_a}^\infty$ are respectively $j_{\mu_a^\Sigma}^\infty$ and $- j_{\lambda_a^\Sigma}^\infty$,
\item the intersection relations (\ref{odotofDlambda}) are given by:
\bea
\int_X j_{\lambda_a}^\infty \wedge j_{D_b}^\infty =  \int_X j_{D_b}^\infty \wedge j_{\lambda_a}^\infty =  - \delta_{ab} = - \int_\Sigma j_{\lambda^{\Sigma}_a}^\infty \wedge j_{\mu^{\Sigma}_b}^\infty \, .
\eea
\end{enumerate}

Let us point out that $j_{D_a}^\infty$ and $j_{A_a}^\infty$ belong to $\Omega^2(X)$ whereas $j_{\lambda_a}^\infty$ is an element of $\Omega^1(X,\partial X)$, the set of relative $1$-forms of $X$, i.e. forms that vanishes on $\partial X$. Of course, $\Omega^1(X,\partial X)$ is canonically embedded into $\Omega^1(X)$.

\vspace{0.5cm}
Before we go to the problem of determining DB cohomology via a Heegaard splitting, let us introduce the following definition:

\begin{defi}\label{defform}\leavevmode

1) A $p$-form of $X_L \cup_f X_R$ can be represented as a couple $(\omega_L,\omega_R) \in \Omega^p(X_L) \times \Omega^p(X_R)$ which fulfills the gluing condition:
\begin{equation}
\label{gluingform}
\omega_R^\Sigma = f_* \, \omega_L^\Sigma \, ,
\end{equation}
where $\omega^\Sigma := i_\Sigma^*(\omega)$. The forms $\omega_L$ and $\omega_R$ will be referred to as the $f$-components of $\omega$.

2) Let $\omega = (\omega_L,\omega_R)$ and $\eta = (\eta_L,\eta_R)$ we two forms of $X_L \cup_f X_R$. Then we set:
\bea
\omega \wedge \eta := (\omega_L \wedge \eta_L , \omega_R \wedge \eta_R) \, .
\eea

3) The de Rham operator acts on a $p$-form $\omega = (\omega_L,\omega_R)$ of $X_L \cup_f X_R$ according to:
\bea
d \omega := (d\omega_L,d\omega_R) \, ,
\eea
$d \omega$ being  a $(p+1)$-form of $X_L \cup_f X_R$.
\end{defi}

Since $d$ commutes with $f_*$, $(d\omega_L,d\omega_R)$ automatically fulfills the gluing condition (\ref{gluingform}) ensuring that the action of $d$ is meaningful. In particular, a $p$-form of $X_L \cup_f X_R$ will be closed if and only if its $f$-components are closed. However, although a $p$-form of $X_L \cup_f X_R$ which is exact has exact $f$-components the converse is not true. For instance, let us consider a closed $2$-form $(F_L,F_R)$ of $X_L \cup_f X_R$. The Poincar\'e-Lefschetz isomorphisms (\ref{PLDisomorphisms}) tell us that $H^2(X) =0$ so that $F_L$ and $F_R$ are necessarily exact. So, if $(F_L,F_R)$ was automatically exact we would have shown that $H^2(M) =0$ for any $M$, which is manifestly untrue.

\subsection{Construction of $H_D^p(X_L \cup_j X_R)$ for $p=1,2,3$}

Let us first recall that a DB $1$-cocycle on a smooth oriented closed $3$-manifold $M$ is nothing but a $U(1)$ gauge potential on $M$. More precisely, if we provide $M$ with a good cover $\mathcal{U}$, then a DB $1$-cocycle of $M$ is given as collection of triples $(A_i,\Lambda_{ij},n_{ijk})$ where the $A_i$'s are $1$-form in the open sets $U_i \in \mathcal{U}$, the $\Lambda_{ij}$'s are $0$-form in the contractible intersections $U_{ij} := U_i \cap U_j$ and the $n_{ijk}$ are integers defined in the triple intersections $U_{ijk} := U_i \cap U_j \cap U_k$, in such a way that all these quantities fulfill the so-called descent equations:
\bea
\label{descentequationM}
\left\{ \begin{aligned}
& A_j - A_i = d \Lambda_{ij} \\
& \Lambda_{jk} + \Lambda_{ik} + \Lambda_{ij} = n_{ijk}
\end{aligned} \right. \, .
\eea
Ehe gluing mapping defining the $U(1)$ principal bundle over $M$ on which the gauge field is defined are given by $g_{ij} := e^{2 i \pi \Lambda_{ij}}$ and the local expressions of the $U(1)$ gauge potential are $a_j := \frac{1}{2i\pi} A_j$. The curvature $2$-form of the DB $1$-cocycle $(A_i,\Lambda_{ij},n_{ijk})$ is  defined locally by the collection of $2$-forms $F_i := d A_i$. Two DB $1$-cocycles $(A_i,\Lambda_{ij},n_{ijk})$ and $(\tilde{A}_i,\tilde{\Lambda}_{ij},\tilde{n}_{ijk})$ are said to be DB equivalent if:
\bea
\label{DBequivalenceM}
\left\{ \begin{aligned}
& \tilde{A}_i - A_i = d q_{i} \\
& \tilde{\Lambda}_{ij} - \Lambda_{ij} = m_{jk} + m_{ik} + m_{ij}
\end{aligned} \right. \, .
\eea
This is nothing but gauge equivalence of $U(1)$ gauge potentials of $M$. The classes of equivalent DV $1$-cocycle of $M$ are called DB $1$-classes and there set is denoted $H_D^1(M)$. This is a $\mathbb{Z}$-module which can be embedded into the following exact sequences:
\bea
\label{DBsequenceM}
\begin{aligned}
0 \rightarrow \frac{\Omega^1(M)}{\Omega^1_{\mathbb{Z}}(M)} \rightarrow \, & H^1_D(M) \rightarrow H^2(M) \rightarrow 0 \\
0 \rightarrow  H^1(M,\mathbb{R}/\mathbb{Z}) \rightarrow \, & H^1_D(M) \rightarrow \Omega^2_\mathbb{Z}(M) \rightarrow 0
\end{aligned} \, ,
\eea
where $H^1(M,\mathbb{R}/\mathbb{Z}) := \Hom\,(H_1(M),\mathbb{R}/\mathbb{Z})$. We refer to the above as the standard construction of $H^1_D(M)$. It turns out that this approach can be generalized to obtain the DB $\mathbb{Z}$-modules $H^p_D(M)$, $p=0,1,2,3$, each of which is embedded into exact sequences analogous to ones above.

The aim of this section is to show that it is possible to recover the standard exacts sequences defining $H^p_D(M)$ by using a Heegaard splitting $M = X_L \cup_j X_R$. The most naive idea is to directly deal with DB classes on $X_L$ and $X_R$ which glue properly via the gluing mapping $f$. However, from the quantum field point of view it is more interesting to try to reconstruct each $H^p_D(M)$ from scratch, that is to say by starting with gauge fields on $X_L$ and $X_R$. As we will see, this approach allows us to identify particular representatives of the DB classes which will be helpful in the context of the CS and BF models.

Before starting the construction, let us recall the exact sequences into which $H^p_D(\Sigma)$, $H^p_D(X)$ and $H^p_D(X,\partial X)$ are embedded \cite{HL01}. First we have the obvious exact sequences:
\bea
\label{pDBsequeforSigma}
\begin{aligned}
0 \rightarrow  \frac{\Omega^p(\Sigma)}{\Omega^p_{\mathbb{Z}}(\Sigma)} \rightarrow \; & H^p_D(\Sigma) \rightarrow H^{p+1}(\Sigma) \rightarrow 0 \\
0 \rightarrow  H^p(\Sigma,\mathbb{R}/\mathbb{Z}) \rightarrow \; & H^p_D(\Sigma) \rightarrow \Omega^{p+1}_{\mathbb{Z}}(\Sigma) \rightarrow 0
\end{aligned} \, ,
\eea
and:
\bea
\label{pDBsequeforX}
\begin{aligned}
0 \rightarrow  \frac{\Omega^p(X)}{\Omega^p_{\mathbb{Z}}(X)} \rightarrow \; & H^p_D(X) \rightarrow H^{p+1}(X) \rightarrow 0 \\
0 \rightarrow  H^p(X,\mathbb{R}/\mathbb{Z}) \rightarrow \; & H^p_D(X) \rightarrow \Omega^{p+1}_{\mathbb{Z}}(X) \rightarrow 0
\end{aligned} \, .
\eea

Moreover, when $p = 1,2,3$, the Poincar\'e-Lefschetz isomorphisms (\ref{PLDisomorphisms}) imply that:
\bea
\label{p123HDX}
H_D^p(X) \simeq \frac{\Omega ^p(X)}{\Omega_\mathbb{Z}^p(X)} \, .
\eea
This means that any $U(1)$ principal bundle over $X$ is trivializable and hence suggests that the gauge fields which have to be used for constructing DB $p$-cocycles of $X_L \cup_j X_R$ are just $p$-forms of $X$. We will now concentrate on these particular cases, leaving aside the quite trivial case $p=0$.

\subsubsection{Construction of $H_D^1(X_L \cup_j X_R)$}

As suggested above, let us consider $(A_L,A_R) \in \Omega^1(X_L) \times \Omega^1(X_R)$. Such a couple defines a $U(1)$ connection (or gauge potential) for the Heegaard splitting $M = X_L \cup_f X_R$ if there exist $\omega_\Sigma \in \Omega^1_{\mathbb{Z}}(\Sigma_R)$ such that:
\bea
\label{gluingforconnection}
A_R^\Sigma - f_* \, A_L^\Sigma = \omega_\Sigma \, .
\eea
The above constraint means that the gauge potentials $A_L$ and $A_R$ match on $\Sigma$ up to a trivial gauge potential, thus ensuring that the DB classes of $A_L$ and $A_R$ exactly glue on $\Sigma$. Note that the set of global gauge transformations on $\Sigma_R$ generates $\Omega^1_{\mathbb{Z}}(\Sigma_R)$, and not just $d{\kern 0.5pt} \Omega^0(\Sigma_R)$.

Since $H^1(\Sigma_R) \simeq H_1(\Sigma_R) \simeq \mathbb{Z}^{2g}$, every $\omega_\Sigma \in \Omega^1_{\mathbb{Z}}(\Sigma_R)$ can be decomposed according to:
\bea
\label{omegaSigmadecomposition}
\omega_\Sigma = \sum_{a=1}^{g} L_\Sigma^a \, j_{\lambda_a^{\Sigma}}^\infty + \sum_{a=1}^{g} M_\Sigma^a \, j_{\mu_a^{\Sigma}}^\infty + d \phi_\Sigma^\infty \, ,
\eea
with $L_\Sigma^a , M_\Sigma^a \in \mathbb{Z}$ and $\phi_\Sigma^\infty \in \Omega^0(\Sigma)$ for every $a=1 , \cdots , g$. We denote by $\vec{L}_\Sigma$ and $\vec{M}_\Sigma$ the corresponding elements of $\mathbb{Z}^g$. When we look at the exact sequences (\ref{pDBsequeforSigma}) we see that the elements of $\Omega^1_{\mathbb{Z}}(\Sigma)$ can be seen as trivial gauge potentials of $\Sigma$ (first exact sequence with $p=1$) but also has curvature of DB $0$-class of $\Sigma$ (second exact sequence with $p=0$). Hence, to each $\omega_\Sigma \in \Omega^1_{\mathbb{Z}}(\Sigma)$ we can associate $\Lambda_\Sigma \in H_D^0(\Sigma)$ such that the curvature of $\Lambda_\Sigma$, $\bar{d} \Lambda_\Sigma$, is $\omega_\Sigma$. Of course, the DB class $\Lambda_\Sigma$ is not unique as the second of the exact sequences (\ref{pDBsequeforSigma}) tells us that two DB $0$-classes have the same curvature if and only if they differ from an element of $H^0(\Sigma,\mathbb{R}/\mathbb{Z}) \simeq \mathbb{R}/\mathbb{Z}$, that is to say from an angle. This suggests the following definition.

\begin{defi}\label{1DBcocycle}\leavevmode

A DB $1$-cocycle of $X_L \cup_f X_R$ is a triplet $\mathcal{A}=(A_L,A_R,\Lambda_\Sigma) \in \Omega^1(X_L) \times \Omega^1(X_L) \times H_D^0(\Sigma_R)$ which fulfills:
\bea
\label{gluing2forA}
A_R^\Sigma - f_* \, A_L^\Sigma = \bar{d} \Lambda_\Sigma \, ,
\eea
\end{defi}

Let us now identify the DB equivalence relation that will yield DB classes. In the standard construction based on a good cover of $M$, this is achieved by identifying DB coboundaries. However, since in the construction we are using handlebodies instead of contractible open sets, the cohomological structure of the construction becomes less obvious. Never the less, since our basic gauge fields are elements of $\Omega^1(X)$ and that we have relation (\ref{p123HDX}), it seems logical to think that the DB ambiguities on the DB $1$-cocycles are generated by elements of $\Omega^1_\mathbb{Z}(X)$.

The Poincar\'e-Lefschetz isomorphisms (\ref{PLDisomorphisms}) imply that we can decompose $\chi_L$ and $\chi_R$ according to:
\bea
\label{DB1ambigdecomp}
\left\{ \begin{aligned}
& \chi_L = \sum_{a=1}^{g} m_L^a j_{D^L_a}^\infty + d q_L \\
& \chi_R = \sum_{a=1}^{g} m_R^a j_{D^R_a}^\infty + d q_R
\end{aligned} \right. \, ,
\eea
with $\vec{m_L}, \vec{m_R} \in \mathbb{Z}^g$ and $(q_L, q_R) \in \Omega^0(X_L) \times \Omega^0(X_R)$. All this yields the following set of definitions in which the symbol $\overline{^{^{\quad ^{}}}}$ denotes the restriction to $\mathbb{R}/\mathbb{Z}$.

\begin{defi}\label{DBequiv&classes}\leavevmode

1) A \textbf{DB ambiguity} for DB $1$-cocycles of $X_L \cup_f X_R$ is a DB $1$-cocycle of the form:
\bea
\Xi_{\vec{m}_L,\vec{m}_R} + Dq \, ,
\eea
where:
\bea
\label{generalambiguity1}
\left\{ \begin{aligned}
& \Xi_{\vec{m}_L,\vec{m}_R} := \left( \sum_{a=1}^{g} m_L^a j_{D_a}^\infty \, , \, \sum_{a=1}^{g} m_R^a j_{D_a}^\infty \, , \, \xi_{\vec{m}_L,\vec{m}_R}^\infty \right) \\
& Dq := \left( d q_L \, , \, d q_R \, , \, \overline{q_R^\Sigma - f_*q_L^\Sigma} \right)
\end{aligned} \right. \, ,
\eea
with $\vec{m}_L, \vec{m}_R \in \mathbb{Z}^g$, $(q_L , q_R) \in \Omega^0(X_L) \times \Omega^0(X_R)$ and $\xi_{\vec{m}_L,\vec{m}_R}^\infty \in H_D^0(\Sigma_R)$ such that:
\bea
\label{xicurvature}
\bar{d} \, \xi_{\vec{m}_L,\vec{m}_R}^\infty = - \sum_{a=1}^{g} (P\vec{m}_L)^a \, j_{\lambda_a^{\Sigma_R}}^\infty + \sum_{a=1}^{g} (\vec{m}_R - Q\vec{m}_L)^a j_{\mu_a^{\Sigma_R}}^\infty - d \left( \sum_{a=1}^{g} m_L^a j_{\psi_a^{\Sigma_R}}^\infty \right)  \, .
\eea

2) Two DB $1$-cocycles $\mathcal{A}$ and $\mathcal{B}$ subordinate to $X_L \cup_f X_R$ are said to be \textbf{DB-equivalent} if they fulfill:
\bea
\label{integralDB1cycle}
\mathcal{B} - \mathcal{A} =  \Xi_{\vec{m}_L,\vec{m}_R} + Dq \, .
\eea

3) The set of DB-equivalent DB $1$-cocycles of $X_L \cup_f X_R$ is a $\mathbb{Z}$-module denoted $H_D^1(X_L \cup_f X_R)$ whose elements are called \textbf{DB classes of degree $1$ of $X_L \cup_f X_R$}.
\end{defi}

\noindent As previously mentioned, the existence of $\xi_{\vec{m}_L,\vec{m}_R}^\infty$ is ensured by the second of the exact sequences (\ref{pDBsequeforSigma}) with $p=0$. and we even know that two such Db $0$-class of $\Sigma_R$ have the same curvature $1$-form $\bar{d} \, \xi_{\vec{m}_L,\vec{m}_R}^\infty$ as given by (\ref{xicurvature}) if and only if they differ by an angle. From now on, $\xi_{\vec{m}_L,\vec{m}_R}^\infty$ will always denote a solution of the curvature equation (\ref{xicurvature}). In fact, we can extend this notation to $\xi_{\vec{x}_L,\vec{x}_R}^\infty$ with $\vec{x}_L, \vec{x}_R \in \mathbb{R}^g$ as long as $\bar{d}\xi_{\vec{x}_L,\vec{x}_R}^\infty \in \Omega^1_\mathbb{Z}(\Sigma_R)$.

The next step is to define the curvature of a DB $1$-cocyle of $X_L \cup_f X_R$.

\begin{defi}\label{CurvatureDB1cocycle}\leavevmode

The \textbf{curvature} of a DB $1$-cocycle $\mathcal{A}=(A_L,A_R,\Lambda_\Sigma)$ of $X_L \cup_f X_R$ is defined as:
\bea
\label{curvatureofA}
\bar{d} \mathcal{A} := (dA_L,dA_R) \in \Omega^2(X_L) \times \Omega^2(X_R) \, .
\eea
It is a closed $2$-form with integral periods of $X_L \cup_f X_R$.

\end{defi}

\noindent The gluing condition (\ref{gluing2forA}) satisfied by $\mathcal{A}$ implies that:
\bea
d A_R^\Sigma - f_* d \, A_L^\Sigma = d (A_R^\Sigma - f_* \, A_L^\Sigma ) = \bar{d} \Lambda_\Sigma = 0 \, .
\eea
Hence, according to Definition \ref{defform}, $\bar{d} \mathcal{A} := (d A_L , d A_R)$ is a $2$-form of $X_L \cup_f X_R$. Moreover it is closed since its representatives in $X_L$ and $X_R$ are exact. In order to check that it has $\bar{d} \mathcal{A}$ we have to identify the closed $2$-cycle of $X_L \cup_f X_R$. As we already noticed, $H_2(X,\partial X)$ is generated by the meridian disks of $X$. Then, it can be checked that the generators of $H_2(X_L \cup_f X_R)$ are, up to some exact contributions, of the form $\mathfrak{S} := \left(\sum_{a=1}^{g} m^a D_a^L , \sum_{a=1}^{g} (Q\vec{m})^a D_a^R \right)$ where $\vec{m} \in \ker P$ and $Q$ is the matrix appearing in (\ref{matrixMf}). Then, it can be shown that:
\bea
\label{periodscurvature}
\oint_{\mathfrak{S}} \bar{d} \mathcal{A} = \sum_{a=1}^{g} (Q\vec{m})^a \oint_{\mu^{\Sigma_R}_a} \bar{d} \Lambda_\Sigma \, ,
\eea
which is an integer as by construction $\bar{d} \Lambda_\Sigma \in \Omega^1_\mathbb{Z}(\Sigma_R)$. Then, it is easy to check that any DB ambiguity $\Xi_{\vec{m}_L,\vec{m}_R} + Dq$ is flat, i.e. it has zero curvature. Hence, two DB equivalent DB $1$-cocycles have the same curvature which yields the exact sequence:
\bea
\label{rightexactsequence1}
H_D^1(X_L \cup_f X_R) \xrightarrow{\bar{d}} \Omega_\mathbb{Z}^2(X_L \cup_f X_R) \rightarrow 0 \, .
\eea
The morphism $\bar{d}$ is surjective,  because any closed $2$-form on $X_L \cup_f X_R$ is represented by an exact couple $(d \alpha_L, d \alpha_R)$ which on its turn give rises to DB $1$-cocycles $(\alpha_L,\alpha_R, \zeta)$.

Let us extend the above exact sequence to the left. To achieve this, we must distinguish inequivalent DB $1$-cocycles which have the same curvature. As the difference of two such DB $1$-cocycles is necessarily flat, we just have to identify inequivalent flat DB $1$-cocycles of $X_L \cup_f X_R$.

Let $\mathcal{A} = (A_L,A_R,\Lambda_\Sigma)$ be a flat DB $1$-cocycle subordinate to $X_L \cup_f X_R$. Its first two components $A_L$ and $A_R$ are then closed $1$-forms. Since by the Poincar\'e-Lefschetz duality theorem we have $H^1(X) \simeq H_2(X, \partial X) \simeq \mathbb{Z}^g$, these forms have the general expression:
\bea
{A_L} = \sum\limits_{a = 1}^g {x_L^a} {\mkern 1mu} j_{D_a^L}^\infty + d{\chi _L} \; \; \; , \; \; \; {A_R} = \sum\limits_{a = 1}^g {x_R^a} {\mkern 1mu} j_{D_a^R}^\infty + d{\chi _R}
 \, ,
\eea
with $\vec{x}_L , \vec{x}_R \in \mathbb{R}^m$ and $(\chi _L,\chi _R) \in \Omega^0(X_L) \times \Omega^0(X_L)$. By subtracting the DB ambiguity $D\chi = (d\chi _L,d\chi _R,\overline{(\chi_R^\Sigma - f_* \chi_L^\Sigma)})$ to $\mathcal{A}$ we obtain a BD equivalent flat cocycle $\mathcal{A}_0 = (a_L,a_R,\lambda_\Sigma)$ such that
\bea
{a_L} = \sum_{a=1}^g x_L^a  j_{D_a^L}^\infty \; \; \; , \; \; \; {a_R} = \sum_{a = 1}^g {x_R^a} j_{D_a^R}^\infty  \, ,
\eea
and which fulfills the gluing condition:
\bea
\bar{d} \lambda_\Sigma = - \sum_{a=1}^{g} (P \vec{x}_L)^a \, j_{\lambda_a^{\Sigma_R}}^\infty + \sum_{a=1}^{g} (\vec{x}_R - Q \vec{x}_L)^a \, j_{\mu_a^{\Sigma_R}}^\infty - d (\sum_{a=1}^{g} x_L^a j_{\psi^{\Sigma_R}_a}^\infty) = \bar{d} \xi_{\vec{x}_L,\vec{x}_R}^\infty \, .
\eea
Hence, we have:
\bea
\mathcal{A}_0 = \left( \sum_{a = 1}^g x_L^a  j_{D_a^L}^\infty \, , \, \sum_{a = 1}^g {x_R^a} j_{D_a^R}^\infty \, , \, \xi_{\vec{x}_L,\vec{x}_R}^\infty \right) \, .
\eea
By construction, $\bar{d} \xi_{\vec{x}_L,\vec{x}_R}^\infty$ has integral periods on $\Sigma_R$, hence it can be written as:
\bea
\label{defvarsigma}
\bar{d} \xi_{\vec{x}_L,\vec{x}_R}^\infty = \sum_{a=1}^{g} L^a \, j_{\lambda_a^{\Sigma_R}}^\infty + \sum_{a=1}^{g} M^a \, j_{\mu_a^{\Sigma_R}}^\infty + d \phi_\Sigma := \bar{d} \zeta_{\vec{L},\vec{M}}^\infty + d \phi_\Sigma \, ,
\eea
for some $\vec{L} , \vec{M} \in \mathbb{Z}^g$, and some $\phi_\Sigma \in \Omega^0(\Sigma_R)$. Consequently,
$\mathcal{A}_0$ is a DB $1$-cocycle of $X_L \cup_f X_R$ if and only if:
\bea
\label{flatconstraints}
\left\{ \begin{gathered}
  {\vec L} =  - P {{\vec x}_L}{\mkern 1mu}  \hfill \\
  {\vec M} = {{\vec x}_R} - Q {{\vec x}_L} \hfill \\
\end{gathered}  \right. \, .
\eea
Moreover, by using the DB ambiguities of type $\Xi_{\vec{m}_L,\vec{m}_R}$, we can reduce $\vec{x}_L$ and $\vec{x}_R$ to be elements of $[0,1[^g$. At the level of classes, this is equivalent to replace $(\vec{x}_L,\vec{x}_R)$ by $(\vec{\theta}_L,\vec{\theta}_R) \in (\mathbb{R}/\mathbb{Z})^g \times (\mathbb{R}/\mathbb{Z})^g$. In doing so, the first of the two constraints (\ref{flatconstraints}) takes the form:
\bea
\label{kerPRconstraint}
P^* \vec{\theta}_L =  \vec{0} \in (\mathbb{R}/\mathbb{Z})^g\, ,
\eea
where $P^*$ is the Pontrjagin dual of the matrix $P$ which appears in (\ref{matrixMf}). More precisely, the matrix $P$ represents a morphism $P:\mathbb{Z}^g \rightarrow \mathbb{Z}^g$ which gives rise to the following exact sequence:
\bea
0 \rightarrow \ker P \rightarrow \mathbb{Z}^g \xrightarrow{P} \mathbb{Z}^g \rightarrow \coker P \rightarrow 0 \, .
\eea
The Pontrjagin dual sequence is then:
\bea
0 \rightarrow \ker P^\star \rightarrow (\mathbb{R}/\mathbb{Z})^g \xrightarrow{P^\star} (\mathbb{R}/\mathbb{Z})^g \rightarrow \coker P^\star \rightarrow 0 \, ,
\eea
which implies that $\ker P^\star \simeq (\coker P)^\star$ and $\coker P^\star \simeq (\ker P)^\star$. Finally, it is not hard to check that $\coker P \simeq H_1(X_L \cup_f X_R)$ so that we have $\ker P^\star \simeq H_1(X_L \cup_f X_R)^\star := H^1(X_L \cup_f X_R,\mathbb{R}/\mathbb{Z})$. Consequently, the constraint (\ref{kerPRconstraint}) is fulfilled if and only if $\vec{\theta} \in H^1(M,\mathbb{R}/\mathbb{Z})$, and we recover the well-known fact that the set of flat DB classes, and thus the set of classes of flat $U(1)$ connections, is isomorphic to $H^1(M,\mathbb{R}/\mathbb{Z})$. This provides us with the left extension of the exact sequence (\ref{rightexactsequence1}) we were looking for, namely:
\bea
\label{exactsequence1DBclasses}
0 \rightarrow H^1(X_L \cup_f X_R,\mathbb{R}/\mathbb{Z}) \rightarrow H_D^1(X_L \cup_f X_R) \rightarrow \Omega_\mathbb{Z}^2(X_L \cup_f X_R) \rightarrow 0 \, .
\eea
This exact sequence describes $H_D^1(X_L \cup_f X_R)$ as a fibration over $\Omega_\mathbb{Z}^2(X_L \cup_f X_R)$ whose translation group along fibers is $H^1(X_L \cup_f X_R,\mathbb{R}/\mathbb{Z})$. Thus, the construction of $H_D^1(X_L \cup_f X_R)$ we have presented here also provides us with representatives of DB classes well-suited to the Heegaard splitting $X_L \cup_f X_R$.

Thus, representatives of flat DB classes, that is to say elements of the translation group $H^1(X_L \cup_f X_R,\mathbb{R}/\mathbb{Z})$, are provided by:
\begin{equation}
\label{flatheegaard}
\mathcal{A}_{\vec{\theta}}^\infty := \left( \sum_{a=1}^{g} \theta^a j_{D_a^L}^\infty \, , \, \sum_{a=1}^{g} (Q \vec{\theta})^a j_{D_a^R}^\infty \, , \, \xi_{\vec{\theta},Q\vec{\theta}}^\infty \right) \, ,
\end{equation}
with $\vec{\theta} \in [0,1[^g$ and $P \vec{\theta} \in \mathbb{Z}^m$. Note that $\bar{d} \xi_{\vec{\theta},Q\vec{\theta}}^\infty = - \sum_{a=1}^{g} (P\vec{\theta})^a j_{\lambda^{\Sigma_R}_a}^\infty - d (\theta^a j_{\psi^{\Sigma_R}_a}^\infty)$ is a curvature on $\Sigma_R$ precisely because $P \vec{\theta} \in \mathbb{Z}^m$. Among the representatives $\mathcal{A}_{\vec{\theta}}^\infty$, we can distinguish, for $\vec{\theta} \neq \vec{0}$, those fulfilling $P \vec{\theta} = \vec{0}$ which correspond to $\vec{\theta} \in F_P$, from those fulfilling $P \vec{\theta} \neq \vec{0}$ and which correspond to $\vec{\theta} \in T_P$. The latter are called \textbf{torsion moves} and the former are called \textbf{free modes}. For later convenience, we denote by $\mathcal{A}_{\vec{\theta}_\tau}^\infty$ and $\mathcal{A}_{\vec{\theta}_f}^\infty$ the torsion origins and free modes, respectively, thus keeping the notation $\mathcal{A}_{\vec{\theta}}^\infty$ for a generic flat DB class with $\vec{\theta} = \vec{\theta}_f + \vec{\theta}_\tau$.

The construction of $H_D^1(X_L \cup_f X_R)$ also provides representatives of origins over the base points of the fibration defined by the exact sequence (\ref{exactsequence1DBclasses}). Indeed, on the fiber over the curvature $\sum_{a=1}^{g} n^a j_{\lambda_a^R}^\infty$ we can consider as origin the DB class of the $1$-cocycle:
\bea
\label{curvheegaard}
\mathcal{A}_{\vec{n}}^\infty := \left( 0 , \sum_{a=1}^{g} n^a j_{A_a^R}^\infty \, , \, \zeta_{- \vec{n}, \vec{0}}^\infty \right) \, ,
\eea
with $\vec{n} \in F_P \simeq \mathbb{Z}^{b_1}$ and where $\zeta_{-\vec{n}, \vec{0}}$ is set from the relation (\ref{defvarsigma}). Thus, the representatives $\mathcal{A}_{\vec{n}}^\infty$ correspond to the free sector of $H_1(X_L \cup_f X_R)$. It is sufficient to add to $\sum_{a=1}^{g} n^a j_{\lambda_a^R}^\infty$ an exact contribution to obtain all the curvatures whose de Rham cohomology class is $\vec{n}$. This corresponds to the non-canonical decomposition $\Omega^2_{\mathbb{Z}}(M) \simeq F^2(M) \times (\Omega^1(M) / \Omega^1_0(M))$. Hence, the representatives that correspond to a change of curvature without change of $\vec{n} \in F_P \simeq \mathbb{Z}^{b_1}$ are of the form:
\bea
\hat{\omega} := \left( \hat{\omega}_L , \hat{\omega}_R , 0 \right) \, ,
\eea
with $\hat{\omega}_R^\Sigma = f_* \hat{\omega}_L^\Sigma$, and thus correspond to global $1$-forms on $M = X_L \cup_f X_R$. At the level of class, we have to pick a $\hat{\omega}$ for each class in $\Omega^1(M)/\Omega^1_0(M)$. This would correspond to fix a particular gauge in this quotient. However, as we will see it in the next section and as we know it from the standard approach \cite{GT2014,MT1}, the contributions of these classes to the CS and BF partition functions will be factorized out and eliminate via a normalization factor in front of the functional integral. This is why, without further details, we admit that we have chosen a representative $\hat{\omega}$ for each class in $\Omega^1(M)/\Omega^1_0(M)$.

Let us make a remark concerning notations. The DB class $\varsigma$ and $\xi$ are related according to:
\bea
\label{relatexi&zeta}
\xi_{\vec{u},\vec{v}}^\infty = \zeta_{-(P\vec{u}),(\vec{v}-Q\vec{u})}^\infty - \sum_{a=1}^{g} u^a j_{\psi^{\Sigma_R}_a}^\infty \, .
\eea
We introduce these DB classes to get rid of the $j_{\psi^{\Sigma_R}_a}^\infty$ terms
which plague the writings of all the representatives and ambiguities.

Let us state the main property of this section which reflect the construction of the exact sequence (\ref{exactsequence1DBclasses}).

\begin{prop}\label{decomp1DB}\leavevmode

The representatives $\mathcal{A}_{\vec{\theta}_\tau}^\infty$, $\mathcal{A}_{\vec{\theta}_f}^\infty$, $\mathcal{A}_{\vec{n}}^\infty$ and $\hat{\omega}$ are independent in the sense that:
\bea
\label{decompuniversal}
H_D^1(X_L \cup_f X_R) \simeq \{ \mathcal{A}_{\vec{n}}^\infty \} \oplus \{ \hat{\omega} \} \oplus \{ \mathcal{A}_{\vec{\theta}_f}^\infty \} \oplus \{ \mathcal{A}_{\vec{\theta}_\tau}^\infty \} \, .
\eea
\end{prop}

\noindent If $\mathcal{A}_{\vec{n}}^\infty = \mathcal{A}_{\vec{\theta}}^\infty$, then $\vec{\theta} = 0$ and hence $\vec{n} = Q\vec{\theta} = 0$. Therefore $\mathcal{A}_{\vec{n}}^\infty = \mathcal{A}_{\vec{\theta}}^\infty = 0$.

\noindent If $\mathcal{A}_{\vec{n}}^\infty = \hat{\omega}$, then $\hat{\omega}_L = 0$ and hence $f_*\hat{\omega}_L^\Sigma = 0 = \hat{\omega}_R^\Sigma$ and so $\sum_{a=1}^{g} n^a j_{\lambda_a^{\Sigma_R}}^\infty = - \hat{\omega}_R^\Sigma = 0$. Therefore, $\vec{n} = 0$ and hence $\mathcal{A}_{\vec{n}}^\infty = \hat{\omega} = 0$.

\noindent If $\mathcal{A}_{\vec{\theta}_f}^\infty = \mathcal{A}_{\vec{\theta}_\tau}^\infty$, then $\vec{\theta}_\tau = \vec{\theta}_f$ so that $P \vec{\theta}_f = \vec{0} = P \vec{\theta}_\tau$. But $P \vec{\theta}_\tau \neq 0$ except for $\vec{\theta}_\tau = \vec{0}$. Therefore $\vec{\theta}_\tau = 0 = \vec{\theta}_f$ and hence $\mathcal{A}_{\vec{\theta}_f}^\infty = \mathcal{A}_{\vec{\theta}_\tau}^\infty = 0$.

\noindent If $\hat{\omega} = \mathcal{A}_{\vec{\theta}_\tau}^\infty$, then $\xi_{\vec{\theta},Q\vec{\theta}}^\infty = 0$ and hence $\bar{d} \xi_{\vec{\theta},Q\vec{\theta}}^\infty = 0$. Then $P \vec{\theta}_\tau = 0$ and therefore $\vec{\theta}_\tau = 0$. Finally $\mathcal{A}_{\vec{\theta}_\tau}^\infty = 0 = \hat{\omega}$.

\noindent If $\hat{\omega} = \mathcal{A}_{\vec{\theta}_f}^\infty$, then $d \hat{\omega}_L = d (\sum_{a=1}^{g} \theta_f^a j_{D_a^L}^\infty) = 0$ and $d \hat{\omega}_R = d (\sum_{a=1}^{g} (Q\vec{\theta}_f)^a j_{D_a^R}^\infty) = 0$. Since $\hat{\omega} \in \Omega^1(M)/\Omega^1_0(M)$, then $\hat{\omega} = 0 = \mathcal{A}_{\vec{\theta}_f}^\infty$.

\noindent QED.

To end this subsection dedicated to $H_D^1(X_L \cup_j X_R)$, let us point out that the exact sequence (\ref{exactsequence1DBclasses}) that we highlighted in our construction is not the one that was used in the standard approach \cite{GT2014,MT2}

\subsubsection{Construction of $H_D^2(X_L \cup_j X_R)$}

The above construction which yielded $H_D^1(X_L \cup_j X_R)$ can also be applied to obtain $H_D^2(X_L \cup_j X_R)$, the second DB cohomology group of $X_L \cup_f X_R$. Although these cohomology classes won't be used in the sequel, let us see how it works in a sake of completeness.

A DB $2$-cocycle of $X_L \cup_f X_R$ is a triplet $\mathcal{F} = (F_L,F_R,A_\Sigma)$, where $(F_L,F_R) \in \Omega^2(X_L) \times \Omega^2(X_L)$ and $A_\Sigma \in H_D^1(\Sigma_R)$, which fulfills the following gluing condition on $\Sigma_R$:
\bea
\label{gluingDBcycle2}
F_R^\Sigma - f_* F_L^\Sigma = \bar{d} A_\Sigma \, ,
\eea
with $\bar{d}$ the canonical injection associated to the exact sequence:
\bea
\label{exacsequSig2}
0 \rightarrow H^1(\Sigma,\mathbb{R}/\mathbb{Z}) \rightarrow H^1_D(\Sigma) \xrightarrow{\bar{d}} \Omega^2_\mathbb{Z}(\Sigma) \rightarrow 0 \, .
\eea
Moreover, as $H^2(X) = 0$, any closed $2$-form on $X$ is exact and hence a DB ambiguity on DB $2$-cocycle is simply of the form $Da = (d a_L, d a_R, \overline{a_R^\Sigma - f_* a_L^\Sigma})$.

The curvature of $\mathcal{F} = (F_L,F_R,A_\Sigma)$ is the $3$-form $\bar{d} \mathcal{F} = (dF_L,dF_R)$. For dimensional reasons, there is no gluing condition to fulfill in this case and $\bar{d} \mathcal{F}$ is necessarily closed. As the only $3$-cycles of $X_L \cup_j X_R$ are $X_L \cup_j X_R$ and its multiples, we can check that $\mathcal{F}$ has integral periods by simply computing its integral a along $M = X_L \cup_f X_R$. This yields:
\bea
\label{periods3curva}
\begin{aligned}
\oint_M  \bar{d} \mathcal{F} = & \int_{X_L} d F_L +  \int_{X_R} d F_R \\
= & \oint_{\Sigma_R} \bar{d} A_\Sigma
\end{aligned} \, ,
\eea
which is an integer since by construction $F_R^\Sigma - f_* F_L^\Sigma = \bar{d} A_\Sigma \in \Omega^2_\mathbb{Z}(\Sigma)$. Obviosuly, a DB ambiguity $Da$ has zero curvature, so that, at the level of classes, we have the exact sequence:
\bea
\label{rightexactsequence2}
H_D^1(X_L \cup_f X_R) \xrightarrow{\bar{d}} \Omega_\mathbb{Z}^2(X_L \cup_f X_R) \rightarrow 0 \, .
\eea
To extend this exact sequence to the left, we consider flat DB $2$-cocycles. A DB $2$-cocycle $(F_L,F_R,A_\Sigma)$ of $X_L \cup_f X_R$ is flat if $d F_L = 0 = d F_R$, and since $H^2(X) \simeq H_1(X,\partial X) = 0$, we deduce that $F_L = d a_L$ and $F_R = d a_R$. Taking into account the DB ambiguities $Da$ already identified and which are associated to $F_L$ and $F_R$, we deduce that a flat DB $2$-cocycle subordinate to $X_L \cup_f X_R$ is always DB-equivalent to a DB $2$-cocycle of form:
\bea
\label{flat2DBnorm}
(0,0,a_\Sigma) \, .
\eea
Hence, to find the DB ambiguities associated to $a_\Sigma$ we can simply consider DB $2$-cocycles of the form $\mathcal{F}_0 = (0,0,a_\Sigma)$. As a DB $2$-cocycle it must fulfill the gluing condition:
\bea
\bar{d} a_\Sigma = 0 - f_*0 = 0 \, ,
\eea
which implies that $a_\Sigma$ is a flat DB $1$-class of $\Sigma_R$, i.e. $a_\Sigma \in H^1(\Sigma,\mathbb{R}/\mathbb{Z}) \simeq (\mathbb{R}/\mathbb{Z})^{g}$. Since the longitudes and meridians of $\Sigma$ are independent generators of $H_1(\Sigma)$, we have that:
\bea
\label{quotientequiv}
H^1(\Sigma,\mathbb{R}/\mathbb{Z}) \simeq \left\{ \sum_{a=1}^{g} \theta^a j_{\lambda_a^\Sigma}^\infty + \sum_{a=1}^{g} \epsilon^a j_{\mu_a^\Sigma}^\infty \right\} \, ,
\eea
with $\vec{\theta} , \vec{\epsilon} \in [0,1[^g$ and therefore, a flat DB class of $\Sigma_R$ as $a_\Sigma$ always admits a representative of the form:
\bea
\label{genericASigma}
\sum_{a=1}^{g} \theta^a j_{\lambda_a^\Sigma}^\infty + \sum_{a=1}^{g} \epsilon^a j_{\mu_a^\Sigma}^\infty \, ,
\eea
with $\vec{\theta} , \vec{\epsilon} \in [0,1[^g$.

A DB $2$-cocycles $(0,0,a_\Sigma)$, with $a_\Sigma$ of the form (\ref{genericASigma}), is a DB ambiguities if and only if $(0,0,a_\Sigma) = (da_L,da_R,\overline{a_R^\Sigma - f_* a_L^\Sigma}) = Da$. This implies that $da_L = 0 = da_R$, and as $H^1(X) \simeq H_2(X,\partial X)$ is generated by the meridian disks $D_a$, this means that:
\bea
a_L = \sum_{a=1}^{g} x_L^a \, j_{D_a^L}^\infty + d \chi_L \; \; \; , \; \; \; a_R = \sum_{a=1}^{g} x_R^a \, j_{D_a^R}^\infty + d \chi_R \, ,
\eea
with $\vec{x}_L, \vec{x}_R \in \mathbb{R}^g$. Hence, the generic form of these $a_\Sigma$ is:
\bea
\label{flatDB2}
\begin{aligned}
a_\Sigma \egzz & \overline{(a_R^\Sigma - f_* a_L^\Sigma) + d(\chi_R^\Sigma - f_* \chi_L^\Sigma)} \egzz \overline{(a_R^\Sigma - f_* a_L^\Sigma)} \\
\egzz & - \sum_{a=1}^{g} (\overline{P\vec{x}_L})^a j_{\lambda_a^{\Sigma_R}}^\infty + \sum_{a=1}^{g} (\overline{\vec{x}_R - Q\vec{x}_L})^a j_{\mu_a^{\Sigma_R}}^\infty
\end{aligned} \, .
\eea
The component of $a_\Sigma$ along $j_{\mu_a^{\Sigma_R}}^\infty$ in (\ref{flatDB2}) can be any element of $(\mathbb{R}/\mathbb{Z})^g$, and hence any DB $2$-cocycle $\left(0,0,\sum_{a=1}^{g} \epsilon^a j_{\mu_a^{\Sigma_R}}^\infty \right)$ is a DB ambiguity. Finally, it is quite obvious that a flat DB $2$-cocycles $(0,0,a_\Sigma)$ is not an ambiguity if and only if $a_\Sigma = \sum_{a=1}^{g} \theta^a j_{\lambda_a^{\Sigma_R}}^\infty$ with $\vec{\theta} \in \overline{\ker P_\mathbb{R}} \simeq (\mathbb{R}/\mathbb{Z})^{b_1} \simeq H^2(X_L \cup_f X_R,\mathbb{R}/\mathbb{Z})$, where $P_\mathbb{R}$ is the canonical extension of $P$ to $\mathbb{R}^g$.

We thus obtain the extension to the left of the exact sequence (\ref{rightexactsequence2}) we were looking for, namely:
\bea
0 \rightarrow H^2(X_L \cup_f X_R,\mathbb{R}/\mathbb{Z}) \rightarrow H_D^2(X_L \cup_f X_R) \xrightarrow{\bar{d}} \Omega_\mathbb{Z}^3(X_L \cup_f X_R) \rightarrow 0 \, .
\eea

Note that we also showed that $(0,0,a_\Sigma)$ is a DB ambiguity if and only if it of the form $Da=(0,0,\overline{a_R^\Sigma - f_* a_L^\Sigma})$, and more generally that any DB ambiguity on DB $2$-cocycles of $X_L \cup_f X_R$ is of the form $Da$. We could call these ambiguities DB $2$-coboundaries of $X_L \cup_f X_R$, thus recovering a cohomological construction of $H_D^2(X_L \cup_f X_R)$.

\subsubsection{Construction of $H_D^3(X_L \cup_j X_R)$}

The last DB space we will need in the sequel is $H_D^3(M)$. In light of the two previous cases, we define a DB $3$-cocycle of $X_L \cup_f X_R$ as a triplet $\mathcal{G} = (\Upsilon_L,\Upsilon_R,\upsilon_\Sigma)$ where $(\Upsilon_L,\Upsilon_R) \in \Omega^3(X_L) \times \Omega^3(X_R)$ and  $\upsilon_\Sigma \in H^2_D(\Sigma)$. Since we have the exact sequence:
\bea
\label{exactDBSigma2}
0 \rightarrow  \frac{\Omega^2(\Sigma)}{\Omega^2_{\mathbb{Z}}(\Sigma)} \rightarrow H^2_D(\Sigma) \rightarrow 0 \, ,
\eea
we can equivalently take $\upsilon_\Sigma \in \Omega^2(\Sigma_R)/\Omega^2_{\mathbb{Z}}(\Sigma_R)$. Moreover, we have $\Omega^2(\Sigma_R)/\Omega^2_{\mathbb{Z}}(\Sigma_R) \simeq \mathbb{R}/\mathbb{Z}$, so that $\upsilon_\Sigma$ is an angle. Let us stress out that there is no gluing condition for DB $3$-cocycle of $X_L \cup_f X_R$ because the restriction to $\Sigma$ of a $3$-form of $X$ is necessarily zero. DB ambiguities of DB $3$-cocycles of $X_L \cup_j X_R$ are generated by $\Omega^3_\mathbb{Z}(X)$. And as $H^3(X) = 0$, this means that a DB ambiguity of DB $3$-cocycle is simply of the form $D \varphi = (d \varphi_L, d \varphi_R, \overline{\varphi_R^\Sigma - f_* \varphi_L^\Sigma})$ with $(\varphi_L, \varphi_R)\in \Omega^2(X_L) \times \Omega^2(X_R)$.

For dimensional reasons, $\Omega^4_\mathbb{Z}(X_L \cup_j X_R) = 0$ and any DB $3$-cocycle of $X_L \cup_j X_R$ is flat. Therefore, the exact sequence:
\bea
H^3_D(X_L \cup_j X_R) \rightarrow \Omega^4_\mathbb{Z}(X_L \cup_j X_R) \rightarrow 0 \, ,
\eea
is trivial. For the same dimensional reasons, $\Upsilon_L$ and $\Upsilon_L$ are closed and hence exact. Therefore, by using a DB ambiguity, any DB $3$-cocycle of $X_L \cup_j X_R$ can be brought to the form $(0,0,\upsilon_\Sigma)$ with $\upsilon_\Sigma \in \mathbb{R}/\mathbb{Z}$. Finally, we have $\rightarrow H^3(X_L \cup_f X_R,\mathbb{R}/\mathbb{Z}) \simeq \mathbb{R}/\mathbb{Z}$ so that we can consider $\upsilon_\Sigma \in \rightarrow H^3(X_L \cup_f X_R,\mathbb{R}/\mathbb{Z})$ which yields the expected exact sequence:
\bea
0 \rightarrow H^3(X_L \cup_f X_R,\mathbb{R}/\mathbb{Z}) \rightarrow H_D^3(X_L \cup_f X_R) \rightarrow 0 \, ,
\eea

The only $3$-cycles of $X_L \cup_j X_R$ are of the form $(nX_L,nX_R)$ and simply represent the $3$-cycles $nM$. The gluing condition for such a $3$-cycle is trivially fulfilled as we have $f(n\Sigma_L) = n f(\Sigma_L) = n\Sigma_R$. This explains why $X_L$ and $X_R$ appears with the same multiplicity in $(nX_L,nX_R)$. The integral of a DB $3$-cocycle $\mathcal{G} = (\Upsilon_L,\Upsilon_R,\upsilon_\Sigma)$ of $X_L \cup_f X_R$ along a $3$-cycle $(nX_L,nX_R)$ is then defined as:
\bea
\label{defintegDB3}
\int_{nM} \mathcal{G} := \int_{nX_L} \Upsilon_L + \int_{nX_R} \Upsilon_L -  \oint_{n \Sigma_R} \upsilon_\Sigma = n \int_{M} \mathcal{G}  \, .
\eea
A integral over $X_L \cup_f X_R$ of a DB ambiguity $D \varphi$ fulfills:
\bea
\begin{aligned}
\int_{M} D \varphi & = \int_{X_L} d \varphi_L + \int_{X_R} d \varphi_R -  \oint_{\Sigma_R} \overline{\varphi_R^\Sigma - f_* \varphi_L^\Sigma} \\
& = \oint_{\Sigma_R} \left\{(\varphi_R^\Sigma - f_* \varphi_R^\Sigma) -  \overline{\varphi_R^\Sigma - f_* \varphi_L^\Sigma} \right\}
\end{aligned} \, ,
\eea
which is an integer by definition of $\overline{\varphi_R^\Sigma - f_* \varphi_L^\Sigma}$. Therefore, integration over $X_L \cup_f X_R$ passes to DB classes as an $\mathbb{R}/\mathbb{Z}$-valued linear mapping. In other words, the integral over $X_L \cup_f X_R$ of a DB $3$-class is defined modulo integers.

As a final remark concerning the DB spaces $H_D^p(X_L \cup_f X_R)$ with $p=1,2,3$, in all these cases there is no DB ambiguity purely coming from the third component of a DB cocycle. DB ambiguities are only associated with the first two components. This is consistent with the point of view we adopt here where the third component in DB cocycles only deal with the gluing condition.

\subsection{DB product and linking form}

There is a pairing:
\bea
\star_D : H_D^1(M) \times H_D^1(M) \rightarrow H_D^3(M) \simeq \mathbb{R}/\mathbb{Z} \, ,
\eea
which is called the DB product of DB classes of degree $1$ of the $3$-manifold $M$. In fact, there is a more general DB product between DB classes of different degrees but this is irrelevant in the present article.

In the standard construction based on a good cover of $M$, the DB product of the DB $1$-cocycle $(A_i,\Lambda_{ij},m_{ijk})$ with the DB $1$-cocycle $(B_i,\Pi_{ij},n_{ijk})$ is given by:
\bea
\left(A_i \wedge d B_i \, , \, \Lambda_{ij} \wedge d B_i \, , \, m_{ijk} \wedge B_k \, , \, m_{ijk} \wedge \Pi_{kl} \, , \, m_{ijk} \wedge n_{klp} \right) \, .
\eea
It is a simple exercise to check that this quintuplet is a DB cocycle. When dealing with a Heegaard splitting $X_L \cup_f X_R$, since there are three geometric components which naturally appear (the two handlebodies $X_L$ and $X_R$ and the surface $\Sigma_R$), the most logical definition of the DB product of two DB $1$-cocycles $\mathcal{A} = (A_L,A_R,\Lambda_\Sigma)$ and $\mathcal{B} = (B_L,B_R,\Pi_\Sigma)$ of $X_L \cup_f X_R$ is the DB $3$-cocycle of $X_L \cup_f X_R$ defined as:
\bea
\label{DBproduct1}
\mathcal{A} \star_D \mathcal{B} = \left(A_L \wedge d B_L , A_R \wedge d B_R , \Lambda_\Sigma \star_D \overline{B_R^\Sigma} \right) \, ,
\eea
with $\overline {^{^{\quad ^{}}}}:\Omega^1(\Sigma_R) \rightarrow H_D^1(\Sigma_R)$ coming from the first exact sequence (\ref{pDBsequeforSigma}) when $p=1$ so that $\Lambda_\Sigma \star_D \overline{B_R^\Sigma} \in H_D^2(\Sigma_R)$, with $\star_D$ the general DB product mentioned above. In fact, since $\overline{B_R^\Sigma}$ is the canonical injection of a form into a DB class we have:
\bea
\Lambda_\Sigma \star_D \overline{B_R^\Sigma} = \overline{B_R^\Sigma} \star_D \Lambda_\Sigma = \overline{B_R^\Sigma \wedge (\bar{d}\Lambda_\Sigma)} = - \overline{(\bar{d}\Lambda_\Sigma) \wedge B_R^\Sigma}\, ,
\eea
so that we can rewrite the previous definition as:
\bea
\label{DBproduct1bis}
\mathcal{A} \star_D \mathcal{B} = \left(A_L \wedge d B_L , A_R \wedge d B_R , \overline{B_R^\Sigma \wedge (\bar{d}\Lambda_\Sigma)} \right) \, ,
\eea

By exchanging $\mathcal{A}$ and $\mathcal{B}$ in this definition, we obtain:
\bea
\label{DBproductexchange}
\mathcal{B} \star_D \mathcal{A} = \left(B_L \wedge d A_L , B_R \wedge d A_R , \overline{A_R^\Sigma \wedge (\bar{d}\Pi_\Sigma)} \right) \, ,
\eea
A simple computation shows that:
\bea
A_R^\Sigma \wedge B_R^\Sigma - f_*(A_L^\Sigma \wedge B_L^\Sigma) = - A_R^\Sigma \wedge (\bar{d} \Pi_\Sigma) - (\bar{d} \Lambda_\Sigma) \wedge B_R^\Sigma - (\bar{d} \Lambda_\Sigma) \wedge (\bar{d} \Pi_\Sigma)
\eea
and hence that:
\bea
\overline{A_R^\Sigma \wedge B_R^\Sigma - f_*(A_L^\Sigma \wedge B_L^\Sigma)} = - \overline{( A_R^\Sigma \wedge (\bar{d} \Pi_\Sigma) + (\bar{d} \Lambda_\Sigma) \wedge B_R^\Sigma )} \, ,
\eea
because as a product of closed forms with integral periods, $(\bar{d} \Lambda_\Sigma) \wedge (\bar{d} \Pi_\Sigma)$ is closed with integral periods too and thus zero in $H_D^2(\Sigma_R)$.

Finally, by adding to the DB $3$-cocycle $\mathcal{B} \star_D \mathcal{A}$ the DB ambiguity:
\bea
\label{ambiguityadded}
\left( d (A_L \wedge B_L) , d (A_R \wedge B_R) , \overline{A_R^\Sigma \wedge B_R^\Sigma - f_*(A_L^\Sigma \wedge B_L^\Sigma)} \right)
\eea
we obtain the DB $3$-cocycle:
\bea
(A_L \wedge d B_L , A_R \wedge d B_R , B_R^\Sigma \wedge(\bar{d} \Lambda_\Sigma)) = \mathcal{A} \star_D \mathcal{B} \, .
\eea
We conclude that up to DB ambiguities, the DB product is commutative on DB $1$-cocycles subordinate to $X_L \cup_f X_R$, as in the standard approach.

For the product $\star_D$ defined on DB cocycles to pass to classes we must check that $\mathcal{A}' \star_D \mathcal{B}' = \mathcal{A} \star_D \mathcal{B}$ when $\mathcal{A}'$ and $\mathcal{B}'$ are DB $1$-cocycles of $X_L \cup_f X_R$ equivalent to $\mathcal{A}$ and $\mathcal{A}$ respectively. Equivalently this means that we must compute the product of a DB $1$-cocycle $\mathcal{A} = (A_L,A_R,\Lambda_\Sigma)$ with a DB ambiguity $\Xi_{\vec{m}_L,\vec{m}_R} + Dq$ and show that the result is a DB $3$-cocycle ambiguity. From definition (\ref{DBproduct1}) we straightforwardly deduce that:
\bea
\label{trivboundaryinproduct}
\mathcal{A} \star_D (\Xi_{\vec{m}_L,\vec{m}_R} + Dq) = \left( 0 , 0 , \Lambda_\Sigma \star_D \overline{(\bar{d} \xi_{\vec{m}_L,\vec{m}_R} + d q_R^\Sigma)} \right) = 0 \, .
\eea
Indeed, by construction $(\bar{d} \xi_{\vec{m}_L,\vec{m}_R} + d q_R^\Sigma) \in \Omega^1_\mathbb{Z}(\Sigma_R)$ so that this $1$-form is zero in the quotient space $\Omega^1(\Sigma_R)/\Omega^1_\mathbb{Z}(\Sigma_R)$ and hence in $H_D^1(\Sigma_R)$. Therefore, $\Lambda_\Sigma \star_D \overline{(\bar{d} \xi_{\vec{m}_L,\vec{m}_R} + d q_R^\Sigma)} = \Lambda_\Sigma \star_D 0 = 0$ in $H_D^3(X_L \cup_f X_R)$ as claimed.

Thanks to the property (\ref{trivboundaryinproduct}), we deduce that the DB product on DB $1$-cocycles of $X_L \cup_f X_R$ goes to DB classes of degree $1$, where it is commutative.

Let us point out that the DB product as defined by formula (\ref{DBproduct1}) is \underline{not} built from the natural pairing in $X$. Indeed, the natural pairing in $X$ involves $H_D^1(X)$ and $H_D^1(X,\partial X)$ \cite{HL01}. But the exterior product $A_L \wedge d B_L$ is not a pairing of this kind since $A_L$ and $B_L$ are both components of elements of $H_D^1(X)$. Furthermore, the pairing defined by the DB product is an example of the more general idea of Pontrjagin duality. However, when dealing with partition functions of the abelian CS and BF theories, the use of Pontrjagin duality can be prevented. It is in the context of expectation values of Wilson loops that the Pontrjagin duality turns out to be more relevant.

\vspace{0.5cm}
The linking form $\Gamma:T_1(M) \rightarrow T_1(M)$ of $M = X_L \cup_f X_R$ can be determine by using the DB product of flat DB $1$-cocycles (or classes). To see this let us first explain how to determine the linking form from a Heegaard splitting. Let $\gamma$ be a torsion $1$-cycle of $X_L \cup_f X_R$. This means that there exist $p \in \mathbb{Z}$ such that $p.\gamma$ is a boundary whereas $\gamma$ is not. Since $H_1(X_L \cup_f X_R)$ is generated by the longitudes of $X_R$, or equivalently of $X_L$, there exist $\vec{l} \in \mathbb{Z}^g$ such that $\gamma = \sum_{a=1}^{g} l^a \lambda^R_a + \partial c_L$. The $1$-cycle $\gamma_{\vec{l}} = \sum_{a=1}^{g} l^a \lambda^R_a$ is also a torsion cycle of order $p$, which means that $p.\gamma_{\vec{l}} = \partial S_{\vec{l}}$, and therefore that:
\bea
\exists \vec{N} \in \mathbb{Z}^g , \; \; \; p \, \vec{l} = P \vec{N} \, .
\eea
Hence, we can write:
\bea
p.\gamma_{\vec{l}} = \sum_{a=1}^{g} \left( P\vec{N} \right)^a \lambda^R_a \, .
\eea
Let us move $\gamma_{\vec{l}}$ to the boundary of $X_R$ by an ambient isotopy and denote by $\gamma_{\vec{l}}^{\Sigma_R}$ the resulting $1$-cycle of $\Sigma_R$. From the relations (\ref{images}), we deduce that:
\bea
\begin{aligned}
p.\gamma_{\vec{l}}^{\Sigma_R} = & \sum_{a=1}^{g} \left( P\vec{N} \right)^a \lambda^{\Sigma_R}_a = f\left( \sum_{a=1}^{g} N^a \mu^{\Sigma_L}_a \right) - \sum_{a=1}^{g} (Q\vec{N})^a \mu^{\Sigma_R}_a \\
= & f\left( \partial (\sum_{a=1}^{g} N^a D^L_a) \right) - \partial \left( \sum_{a=1}^{g} (Q\vec{N})^a D^R_a \right) \, .
\end{aligned} \, .
\eea
The $2$-chain $\sum_{a=1}^{g} N^a D^L_a$ lies in $X_L$ and the $2$-chain $\left( \sum_{a=1}^{g} (Q\vec{N})^a D^R_a \right)$ lies in $X_R$, and therefore $\gamma_{\vec{l}}$ is only intersecting the latter. The self-linking of $\gamma_{\vec{l}}$ is defined by:
\bea
lk(\gamma_{\vec{l}}^{\Sigma_R},\gamma_{\vec{l}}) = \frac{1}{p} \left( - \sum_{a=1}^{g} (Q\vec{N})^a D^R_a \right) \odot \gamma_{\vec{l}} \, ,
\eea
which yields:
\bea
\begin{aligned}
lk(\gamma_{\vec{l}}^{\Sigma_R},\gamma_{\vec{l}}) = & - \frac{1}{p} \left( \sum_{a=1}^{g} (Q\vec{N})^a D^R_a \right) \odot \left( \sum_{b=1}^{g} \frac{1}{p} (P\vec{N})^b \lambda^R_b \right) \\
= & \sum_{a,b=1}^{g} (\frac{1}{p} Q\vec{N})^a \delta_{ab} (\frac{1}{p} P\vec{N})^b \\
:= & \left\langle Q \left(\frac{\vec{N}}{p}\right) , P \left(\frac{\vec{N}}{p}\right) \right\rangle
\end{aligned} \, .
\eea
If we set $\vec{\theta}_\tau = \vec{N}/p$, we obtain the following expression of the linking form of $X_L \cup_f X_R$:
\bea
\Gamma(\vec{\theta}_\tau , \vec{\theta}_\tau) \egzz \left\langle Q \vec{\theta}_\tau , P \vec{\theta}_\tau \right\rangle
\eea
This construction straightforwardly extend to two different torsion cycles, thus yielding:
\bea
\label{linkingformM}
\begin{aligned}
\Gamma(\vec{\theta}_\tau , \vec{\vartheta}_\tau) \egzz & \left\langle Q \vec{\theta}_\tau , P \vec{\vartheta}_\tau \right\rangle = \left\langle \vec{\theta}_\tau , Q^\dag P \vec{\vartheta}_\tau \right\rangle = \left\langle \vec{\theta}_\tau , P^\dag Q \vec{\vartheta}_\tau \right\rangle = \left\langle P \vec{\theta}_\tau , Q \vec{\vartheta}_\tau \right\rangle \\
\egzz & \Gamma(\vec{\vartheta}_\tau , \vec{\theta}_\tau)
\end{aligned} \, ,
\eea
where we used the first of the properties (\ref{mainpropPQRS}).

Now we can state the following property.

\begin{prop}\label{linkformDB}\leavevmode

The torsion moves of $H_D^1(X_L \cup_f X_R)$ determine the linking form of $X_L \cup_f X_R$ according to the relation:
\bea
\Gamma(\vec{\theta}_\tau,\vec{\vartheta}_\tau) \egzz - \int_{M} \mathcal{A}_{\vec{\theta}_\tau}^\infty \star_D \mathcal{A}_{\vec{\vartheta}_\tau}^\infty \egzz \left\langle Q \vec{\theta}_\tau , P \vec{\theta}_\tau \right\rangle \, .
\eea
\end{prop}

\noindent Indeed, as $d j_{D_a}^\infty = 0$ for $a=1,\cdots,g$, the DB product $\mathcal{A}_{\vec{\theta}_\tau}^\infty \star_D \mathcal{A}_{\vec{\vartheta}_\tau}^\infty$ is reduced to
\bea
\left( 0,0, \sum_{a=1}^{g} (Q \vec{\theta}_\tau)^a j_{\mu^{\Sigma_R}_a}^\infty \wedge \sum_{b=1}^{g} \left( (-P \vec{\theta}_\tau)^b j_{\lambda^{\Sigma_R}_b}^\infty + d (\theta^b j_{\psi^{\Sigma_R}_b}^\infty)  \right) \right) \, ,
\eea
and hence, after an integration by part which eliminates the $j_{\psi^{\Sigma_R}_b}^\infty$ contributions, we have:
\bea
\label{linkingformfromDB}
\begin{aligned}
\int_{M} \mathcal{A}_{\vec{\theta}_\tau}^\infty \star_D \mathcal{A}_{\vec{\vartheta}_\tau}^\infty = & - \oint_{\Sigma_R} \left( \sum_{a=1}^{g} (Q \vec{\theta}_\tau)^a j_{\mu^{\Sigma_R}_a}^\infty \wedge \sum_{b=1}^{g} (-P \vec{\vartheta}_\tau)^b j_{\lambda^{\Sigma_R}_b}^\infty \right) \\
= & - \sum_{a,b=1}^{g}  (Q \vec{\theta}_\tau)^a \delta_{ab} (P \vec{\vartheta}_\tau)^b \\
= & - \Gamma(\vec{\theta}_\tau,\vec{\vartheta}_\tau)
\end{aligned} \, ,
\eea
which is the announced result. The bilinear and commutative nature of the DB product is equivalent to the bilinear and symmetric character of the linking form.

\section{Partition Functions of the $U(1)$ CS and BF theories}

As already mentioned in the introduction of this article, the $U(1)$ CS and BF theories have been completely solved by using DB cohomology of $M$ \cite{GT2008,GT2013,GT2014,MT1,MT2}. The procedure was to express the action of each of these theories, then to inject it into the functional integral defining the partition functions and then to extract from these formal integral the finite part which yield the RT and TV invariants. We would like to revisit this procedure from the point of view of the Heegaard splitting construction above.

\subsection{CS and BF actions}

As already explained in the series of original articles about the use of DB cohomology in the context of $U(1)$ CS and BF theories, the actions for these theories are respectively:
\bea
\label{CSaction}
S_{CS,k} = k \int_M \mathcal{A} \star_D \mathcal{A} \, , \\
\label{BFaction}
S_{BF,k} = k \int_M \mathcal{A} \star_D \mathcal{B} \, ,
\eea
where $\mathcal{A},\mathcal{B} \in H_D^1(M)$, $\star_D$ being the DB product and $k$ is a coupling constant. In order to write these actions in the context of Heegaard splitting, we first have to express the DB product of DB classes in term of the representatives introduced in the previous section. So, let us consider the decomposition:
\begin{equation}
\label{A}
\mathcal{A} = \mathcal{A}_{\vec{n}}^\infty + \hat{\omega} + \mathcal{A}_{\vec{\theta}}^\infty = \mathcal{A}_{\vec{n}}^\infty + \hat{\omega} + \mathcal{A}_{\vec{\theta}_f}^\infty + \mathcal{A}_{\vec{\theta}_\tau}^\infty \, ,
\end{equation}
hence with $\vec{\theta} = \vec{\theta}_f + \vec{\theta}_\tau$. Let us recall that in these expressions we are dealing with our set of particular representatives. For instance $\vec{\theta} \in [0,1[^g$ such that $P \vec{\theta} \in \mathbb{Z}^g$. In fact $\vec{\theta}$ represents an element of $\coker P^* \simeq H^1(M,\mathbb{R}/\mathbb{Z}) \simeq (\mathbb{R}/Z)^{b_1} \times \mathbb{Z}_{p_1} \times \cdots \times \mathbb{Z}_{p_N}$ so that $\vec{\theta}_f$ represents an element of $T_P \simeq (\mathbb{R}/Z)^{b_1}$ and $\vec{\theta}_\tau$ an element of $T_P \simeq T_1 \simeq \mathbb{Z}_{p_1} \times \cdots \times \mathbb{Z}_{p_N}$.

Before doing this, let us denote $\mathcal{A} = (A_L,A_R,\Lambda_\Sigma)$ and $\mathcal{B} = (B_L,B_R,\Pi_\Sigma)$ DB $1$-cocycles of $X_L \cup_f X_R$ representing in the DB classes appearing in the actions (\ref{CSaction}) and (\ref{BFaction}). Then, with respect to the components of $\mathcal{A}$ and $\mathcal{B}$, the CS and BF action respectively read:
\bea
\label{actionsHS}
S_{CS,k} &\egzz& k \left\{ \int_{X_L} A_L \wedge d A_L +  \int_{X_R} A_R \wedge d A_R - \int_{\Sigma_R} \overline{A_R^\Sigma \wedge (\bar{d} \Lambda_\Sigma)} \right\} \, , \\
S_{BF,k} &\egzz& k \left\{ \int_{X_L} A_L \wedge d B_L +  \int_{X_R} A_R \wedge d B_R - \int_{\Sigma_R} \overline{B_R^\Sigma \wedge (\bar{d} \Lambda_\Sigma)} \right\} \, ,
\eea
where $k$ is a coupling constant which must be an integer as the expressions in braces are defined in $\mathbb{R}/\mathbb{Z}$.

Let us now write the CS and BF actions in term of the particular representatives $\mathcal{A} = \mathcal{A}_{\vec{\theta}}^\infty + \mathcal{A}_{\vec{m}}^\infty + \hat{\omega}$ and $\mathcal{B} = \mathcal{B}_{\vec{\vartheta}}^\infty + \mathcal{B}_{\vec{n}}^\infty + \hat{\eta}$ introduced in the previous section, starting with the CS action. With respect to the expressions (\ref{actionsHS}), we have:
\bea
\left\{ \begin{aligned}
& A_L = \sum_{a=1}^{g} \theta^a j_{D^L_a}^\infty + \hat{\omega}_L   \\
& A_R =  \sum_{a=1}^{g} m^a j_{A^R_a}^\infty + \sum_{a=1}^{g} (Q \vec{\theta})^a j_{D^R_a}^\infty +  \hat{\omega}_R  \\
& \bar{d} \Lambda_\Sigma =  - \sum_{a=1}^{g} (\vec{m} + P \vec{\theta})^a j_{\lambda_a^{\Sigma_R}}^\infty
\end{aligned} \right. \, ,
\eea
with $\vec{\theta} \in [0,1[^g$ and $P \vec{\theta} \in \mathbb{Z}^g$, and $P^\dag \vec{m} = 0$. Therefore we have:
\bea
A_L \wedge d A_L = \left( \sum_{a=1}^{g} \theta^a j_{D_a^L}^\infty + \hat{\omega}_L \right) \wedge d \hat{\omega}_L  \, ,
\eea
since $d{\kern 0.3pt} j_{D_a}^\infty = 0$. Similarly we have:
\bea
A_R \wedge d A_R = \left( \sum_{a=1}^{g} m^a j_{A_a^R}^\infty +\sum_{a=1}^{g} (Q \vec{\theta})^a j_{D_a^R}^\infty +  \hat{\omega}_R \right) \wedge \left( \sum_{a=1}^{g} m^a j_{\lambda_a^R}^\infty + d\hat{\omega}_R  \right) \, ,
\eea
since $d{\kern 0.3pt} j_{A_a}^\infty = j_{\lambda_a}^\infty$, as well as:
\bea
A_R^\Sigma \wedge (\bar{d} \Lambda_\Sigma) =  \left( \sum_{a=1}^{g} m^a j_{\lambda^{\Sigma_R}_a}^\infty - \sum_{a=1}^{g} (Q \vec{\theta})^a j_{\mu^{\Sigma_R}_a}^\infty -  \hat{\omega}_R^\Sigma \right) \wedge \sum_{a=1}^{g} (\vec{m} + P \vec{\theta})^a j_{\lambda_a^{\Sigma_R}}^\infty \, .
\eea
After few integrations by parts, we obtain:
\bea
S_{CS,k} \egzz \! \! \! &k& \! \! \! \! \! \left( \int_{X_L} \hat{\omega}_L \wedge d \hat{\omega}_L
  + \int_{X_R} \hat{\omega}_R \wedge d \hat{\omega}_R \right) + 2 k \sum_{a=1}^{g} m^a \int_{X_R} \hat{\omega}_R \wedge j_{\lambda_a^R}^\infty + \nonumber \\
   &-& \! \! \! \! 2k \sum_{a,b=1}^{g} m^a \delta_{ab} \, \theta_f^b - k \sum_{a,b =1}^{g} (Q \vec{\theta}_\tau)^a \delta_{ab} (P \vec{\theta}_\tau)^b   \, ,
\eea
with $\vec{\theta} = \vec{\theta}_f + \vec{\theta}_\tau$ such that $P \vec{\theta}_f = 0$. Since we saw that:
\bea
\label{recalllinkformtors}
\sum_{a,b =1}^{g} (Q \vec{\theta}_\tau)^a \delta_{ab} (P \vec{\theta}_\tau)^b = \left\langle Q \vec{\theta}_\tau , P \vec{\theta}_\tau \right\rangle \egzz \Gamma(\vec{\theta}_\tau,\vec{\theta}_\tau) \, ,
\eea
with $\Gamma: T_1(M) \rightarrow T_1(M)$ the linking form of $M = X_L \cup_f X_R$, we finally obtain the following expression for the CS action:
\bea
\label{CSfinal}
S_{CS,k} \egzz \! \! \! &k& \! \! \! \! \! \left( \int_{X_L} \hat{\omega}_L \wedge d \hat{\omega}_L
  + \int_{X_R} \hat{\omega}_R \wedge d \hat{\omega}_R \right) + 2 k \sum_{a =1}^{g} m^a \int_{X_R} \hat{\omega}_R \wedge j_{\lambda_a^R}^\infty \\
   &-& \! \! \! \! 2k \left\langle \vec{\theta}_f , \vec{m}  \right\rangle - k \, \Gamma(\vec{\theta}_\tau , \vec{\theta}_\tau) \, , \nonumber
\eea

To determine $S_{BF}$ we proceed in the same way. We introduce the fields:
\bea
\left\{ \begin{aligned}
& B_L = \sum_{a=1}^{g} \vartheta^a j_{D^L_a}^\infty +  \hat{\omega}_L \\
& B_R = \sum_{a=1}^{g} n^a j_{A^R_a}^\infty + \sum_{a=1}^{g} (Q \vec{\vartheta})^a j_{D^R_a}^\infty +  \hat{\omega}_R \, , \\
& \bar{d} \Pi_\Sigma = - \sum_{a=1}^{g} (\vec{n} + P \vec{\vartheta})^a j_{\lambda_a^{\Sigma_R}}^\infty
\end{aligned} \right. \, ,
\eea
with $\vec{\vartheta} \in [0,1[^g$ and $P \vec{\vartheta} \in \mathbb{Z}^g$. A computation totally similar to the one that yields the CS action gives:
\bea
\label{BFfinal}
S_{BF,k} \egzz \! \! \! &k& \! \! \! \! \! \left( \int_{X_L} \hat{\omega}_L \wedge d \hat{\eta}_L + \int_{X_R} \hat{\omega}_R \wedge d \hat{\eta}_R \right) + \\
&\;& \; \; \; \; \; \; + \, k \sum_{a =1}^{g} n^a \int_{X_R} \hat{\omega}_R \wedge j_{\lambda_a^R}^\infty + k \sum_{a =1}^{g} m^a \int_{X_R} \hat{\eta}_R \wedge j_{\lambda_a^R}^\infty + \nonumber \\
&\;& \; \; \; \; \; \; \; \; \; \; \; \; \; \; \; \; \; \; - \, k \left\langle \vec{\theta}_f , \vec{n} \right\rangle - k \left\langle \vec{\vartheta}_f , \vec{m} \right\rangle - k \, \Gamma (\vec{\theta}_\tau, \vec{\vartheta}_\tau)  \, , \nonumber
\eea
with $\vec{\theta} = \vec{\theta}_f + \vec{\theta}_\tau$ and $\vec{\vartheta} = \vec{\vartheta}_f + \vec{\vartheta}_\tau$ such that $P\vec{\theta}_f = 0 = P\vec{\vartheta}_f$. When $\mathcal{B} = \mathcal{A}$, the expression (\ref{BFfinal}) yields the CS action (\ref{CSfinal}), as it should be.

Let us point out that the determination of the CS and BF actions indirectly provides us with a set of properties concerning the DB product of the representatives used to decompose a DB class. Namely we have proven that:
\bea
\left\{ \begin{aligned}
& \int_M \mathcal{A}_{\vec{m}}^\infty \star_D \mathcal{A}_{\vec{\theta}_\tau}^\infty \egzz - \left\langle \vec{m} , \vec{\theta} \right\rangle \\
& \int_M \mathcal{A}_{\vec{m}}^\infty \star_D \hat{\omega} \egzz \sum_{a =1}^{g} m^a \oint_{\lambda_a^R} \hat{\omega}_R \wedge  \\
& \int_M \hat{\omega} \star_D \hat{\eta} \egzz \int_{X_L} \hat{\omega}_L \wedge d \hat{\eta}_L + \int_{X_R} \hat{\omega}_R \wedge d \hat{\eta}_R \\
& \int_{M} \mathcal{A}_{\vec{\theta}_\tau}^\infty \star_D \mathcal{A}_{\vec{\vartheta}_\tau}^\infty \egzz - \Gamma(\vec{\theta}_\tau,\vec{\vartheta}_\tau)
\end{aligned} \right. \, ,
\eea
all other DB products being zero (mod. $\mathbb{Z}$). Let us note that the last line in the set of above equalities is nothing but Property \ref{linkformDB}.

Now that we have these expressions for the CS and BF actions we can go to the determination of the corresponding partition functions.

\subsection{Partition Functions from functional integration}

Once the classical actions have been defined, the definition of the corresponding partition functions are given through formal functional integrals. More precisely, we set:
\bea
\label{CSpartitionfunc}
\mathcal{Z}_{CS,k} &=& \frac{1}{\mathcal{N}_{CS,k}} \int D\!A \; e^{2 i \pi S_{CS}} \, , \\
\label{BFpartitionfunc}
\mathcal{Z}_{BF,k} &=& \frac{1}{\mathcal{N}_{BF,k}} \int D\!A  \, D\!B\; e^{2 i \pi S_{BF}} \, .
\eea
In the standard approach like the one developed in \cite{GT2014} and \cite{MT2}, the functional integration is performed on DB classes. This corresponds to the idea of directly dealing with gauge classes of gauge potentials, without the need of a gauge fixing procedure. Strictly speaking, we used representatives of DB classes in the expressions of the CS and BF actions given above. In that sense, we also get ride of the gauge invariance, but in a different way. In our Heegaard splitting context, the functional ``measure" $\int DA$ must be written according to the representatives, which means as:
\bea
\label{functmeasure}
\int D\!A := \sum_{\vec{n} \in F_P} \, \sum_{\vec{\theta}_\tau \in T_P} \, \oint (d^{b_1}{\vec{\theta}_f}) \, \int (D\hat{\omega}_L) \, \int (D\hat{\omega}_R) \,
\eea
This expression for the functional measure is meaningful thanks to Property IV. In this expression of the functional measure the first three ``integrals" are finite, so that only the last two are truly functional integrals. As in the standard approach \cite{GT2014,MT1}, we will see that, in CS as well as in BF, these functional integrals are eventually factorized out and simplified thanks to the normalization factors $\mathcal{N}_{CS,k}$ or $\mathcal{N}_{BF,k}$.

The $U(1)$ CS functional measure is $D\!A \; e^{2i \pi S_{CS,k}}$. The exponential appearing in this measure is referred to as the CS measure density, and with respect to the detailed expression (\ref{CSfinal}) of the CS action, this density takes the form:
\bea
\label{integrandCSmeasure}
e^{ 2i \pi k \left( \int_{X_L} \hat{\omega}_L \wedge d \hat{\omega}_L
+ \int_{X_R} \hat{\omega}_R \wedge d \hat{\omega}_R  + 2 \sum_{a =1}^{g} m^a \int_{X_R} \hat{\omega}_R \wedge j_{\lambda_a^R}^\infty -  2 \left\langle \vec{\theta} , \vec{m}  \right\rangle -  \, \Gamma( \vec{\theta},  \vec{\theta}) \right) } \, .
\eea

The CS functional measure have \textbf{zero modes}. To see this, let us consider the shift:
\bea
\label{zeromodes}
\vec{\theta} \rightarrow \vec{\theta} + \frac{\vec{u}}{2k} \, ,
\eea
with $\vec{u} \in \ker P$. This shift implies the following changes:
\bea
\left\{ \begin{aligned}
-2k \left\langle \vec{\theta} , \vec{m}  \right\rangle \rightarrow & \left\langle \vec{\theta} , \vec{m}  \right\rangle - \left\langle \vec{u} , \vec{m}  \right\rangle \egzz \left\langle \vec{\theta} , \vec{m}  \right\rangle \\
\left\langle Q \vec{\theta} , P \vec{\theta}  \right\rangle \rightarrow & \left\langle Q \vec{\theta} , P \vec{\theta}  \right\rangle
\end{aligned} \right. \, .
\eea
We used the fact that $P \vec{u} =0$ and the property $Q^\dag P = P^\dag Q$ in order to obtain the second line above. Hence, the measure density is invariant under the shift (\ref{zeromodes}). To make contact with zero modes, let us remark that the shift (\ref{zeromodes}) corresponds to a shift by:
\bea
\label{zeromodegeom}
\left( \sum_{a=1}^{g} \frac{u^a}{2k} j_{D^L_a}^\infty , \sum_{a=1}^{g} (Q\frac{\vec{u}}{2k})^a j_{D^R_a}^\infty \right) \, ,
\eea
which is obviously a free mode $\mathcal{A}_{\vec{\theta}_f}^\infty$ with $\vec{\theta}_f = \vec{u}/2k$ since $P \vec{u} =0$. Hence zero modes are particular free modes, namely those such that $\vec{\theta}_f \in (\mathbb{Z}_{2k})^g$. When $\vec{u}$ is a multiple of $2k$ the corresponding zero mode becomes a DB ambiguity and thus vanishes. This is sometimes referred to as the $2k$ periodicity of the $U(1)$ CS theory.

Let us return to our aim, the determination of the CS and BF partition functions. First, we perform the finite integration over the free modes $\vec{\theta}_f$. This finite integration yields:
\bea
\oint_{(\mathbb{R}/\mathbb{Z})^{b_1}} d^{b_1}\vec{\theta}_f \; e^{- 2i \pi (2k) \left\langle \vec{\theta}_f , \vec{m} \right\rangle } = \delta_{\vec{m},\vec{0}} \, .
\eea
Now, let us perform the finite integration over $\vec{m} \in F_P \simeq \mathbb{Z}^{b_1}$. Taking into account the previous result, we obtain:
\bea
\sum_{\vec{m} \in F_P} \delta_{\vec{m},\vec{0}} \; e^{2i\pi (2k) \sum_{a =1}^{g} m^a \int_{X_R} \hat{\omega}_R \wedge j_{\lambda_a^R}^\infty} = 1 \, .
\eea
Hence, the partition function $\mathcal{Z}_{CS,k}$ can be simply written as:
\bea
\sum_{\vec{\theta}_\tau \in T_P} \int (D\hat{\omega}_L) \, \int (D\hat{\omega}_R) \; e^{ 2i \pi k \left( \int_{X_L} \hat{\omega}_L \wedge d \hat{\omega}_L
+ \int_{X_R} \hat{\omega}_R \wedge d \hat{\omega}_R \right)} e^{-2i \pi k \Gamma( \vec{\theta}_\tau,  \vec{\theta}_\tau)} \, .
\eea
The integrals over $\hat{\omega}_L$ and $\hat{\omega}_R$ are decoupled from the sum over the torsion. The former is still a functional integral whereas the later is a finite sum. Then, as already mentioned, we can eliminate the infinite dimensional integrals by setting:
\bea
\mathcal{N}_{CS,k} &=& \int (D\hat{\omega}_L) \, \int (D\hat{\omega}_R) \; e^{ 2i \pi k \left( \int_{X_L} \hat{\omega}_L \wedge d \hat{\omega}_L
+ \int_{X_R} \hat{\omega}_R \wedge d \hat{\omega}_R \right)} \\
&=& \int (D\hat{\omega}) \; e^{ 2i \pi k \left( \int_{M} \hat{\omega} \wedge d \hat{\omega} \right)} \, ,
\eea
so that we finally obtain:
\bea
\mathcal{Z}_{CS,k} = \sum_{\vec{\theta}_\tau \in T_P} e^{-2i k \pi \Gamma( \vec{\theta}_\tau,  \vec{\theta}_\tau)} \, .
\eea
This is the well known result of the standard approach \cite{GT2013}.

Let us point out that it might be surprising to factorize out and then eliminate the infinite dimensional contribution $\mathcal{N}_{CS,k}$ to the functional integration. We can give two related justifications of this. Firstly, the partition function $\mathcal{Z}_{CS,k}$ thus obtained coincides, after applying some Gauss reciprocity formula, with the RT abelian invariant. Secondly, the normalization factor $\mathcal{N}_{CS,k}$ is somewhat universal in the sense that it is the only contribution to the functional integral in the case where $M = S^3$. So, to eliminate this contribution via $\mathcal{N}_{CS,k}$ ensures us that we have $\mathcal{Z}_{CS,k}(S^3) = 1$. In fact we also have $\mathcal{Z}_{CS,k}(S^1 \times S^2) = 1$, so it is not totally true that our choice of normalization is fundamentally associated with $\mathcal{Z}_{CS,k}(S^3) = 1$, but at least it is consistent with it. This ambiguity on the interpretation of the normalization in the abelian CS case is also present in the abelian RT construction \cite{MT1}.

To determine $\mathcal{Z}_{BF,k}$ we just have to repeat the previous procedure. The only difference is that every thing is doubled in the functional BF measure:
\bea
 \sum_{\vec{m},\vec{n} \in F_P} \; \sum_{\vec{\theta}_\tau,\vec{\vartheta}_\tau \in T_P} \; \left( (d^{b_1}\vec{\theta}_f) (d^{b_1}\vec{\vartheta}_f) \right)
\; \left((D\hat{\omega}_L) (D\hat{\eta}_L) (D\hat{\omega}_R) (D\hat{\eta}_R)\right) \, e^{2i\pi S_{BF,k}} \, .
\eea
The zero modes associated with the BF functional measure are now associated to the shifts:
\bea
\left\{ \begin{aligned}
\vec{\theta} \rightarrow \vec{\theta} + \frac{\vec{m_1}}{k} \\
\vec{\vartheta} \rightarrow \vec{\vartheta} + \frac{\vec{m_2}}{k}
\end{aligned} \right. \, ,
\eea
with $\vec{m_1},\vec{m_2} \in \ker P$. It is easy to check that $e^{2i\pi S_{BF}}$ is invariant under such a shift. The BF theory is $k$-periodic.

To get $\mathcal{Z}_{CS,k}$, we first perform the integrations over the free modes $\vec{\theta}_f$ and $\vec{\vartheta}_f$, thus obtaining:
\bea
\oint \oint (d^{b_1}\vec{\theta}_f) (d^{b_1}\vec{\vartheta}_f)  \; e^{- 2i \pi (2k) ( \left\langle \vec{n} , \vec{\theta}_f \right\rangle + \left\langle \vec{m} , \vec{\vartheta}_f \right\rangle ) } = \delta_{\vec{m},\vec{0}} \, \delta_{\vec{n},\vec{0}} \, .
\eea
Then we perform the sum over $\vec{m}$ and $\vec{n}$ which, thanks to the previous delta contributions, gives $1$. As in the CS case, the remaining integrals decouple from the remaining sums over $\vec{\theta}_\tau$ and $\vec{\vartheta}_\tau$, so that if we set:
\bea
\mathcal{N}_{BF,k} = \int \left((D\hat{\omega}_L) (D\hat{\eta}_L) (D\hat{\omega}_R) (D\hat{\eta}_R)\right) \; e^{ 2i \pi k \left( \int_{X_L} \hat{\omega}_L \wedge d \hat{\eta}_L
+ \int_{X_R} \hat{\omega}_R \wedge d \hat{\eta}_R \right)} \, ,
\eea
then the BF partition function is:
\bea
\mathcal{Z}_{BF,k} = \sum_{\vec{\theta}_\tau \in T_P} \sum_{\vec{\vartheta}_\tau \in T_P} \, e^{ -2i k \pi \Gamma(\vec{\theta}_\tau,\vec{\vartheta}_\tau) } \, ,
\eea
this expression coinciding with the standard expression of the $U(1)$ BF partition function \cite{MT1}.

\section{Conclusion}

We have shown how the use of the Heegaard splitting construction for smooth closed oriented $3$-manifold is natural in the context of $U(1)$ CS and BF theories. Of course, it is still DB cohomology which is behind the scene, like in the standard approach. However, we rather used specific representatives of DB classes instead of DB classes themselves as what is usual done in the standard approach. This way to do seems closer to the Quantum Field Theory technics (gauge fixing procedure to fix a representative in each gauge field class) than the standard approach which deals with classes. It also shows that we eventually perform only finite integrals on ``natural" parameters. Since we also know that the non-abelian classical CS invariant takes a particularly nice form when using Heegaard splitting, we can wonder whether the computation of the non-abelian CS partition function could be done with the help of a Heegaard splitting and in the functional integral context.

Of course, the next step would be show how the expectation values of Wilson loops can also be performed in the Heegaard splitting as an extension of what has been done in this article. In order to achieve this, it could be interesting to discuss the role of Pontrjagin duality as it is the natural mathematical framework in which de Rham-Federer currents representing cycles (loops) appear as DB quantities \cite{HLZ03,HL01}.

As a lat remark, let us point out that the Heegaard splitting yield gauge class representatives which are connected with the exact sequence (\ref{exactsequence1DBclasses}). However, in the standard approach it is the exact sequence:
\bea
0 \rightarrow \frac{\Omega^1(M)}{\Omega^1_\mathbb{Z}(M)} \rightarrow H^1_D(M) \rightarrow H^2(M) \rightarrow 0 \, ,
\eea
which was used. In some occasions, we wondered whether it could be possible to use (\ref{exactsequence1DBclasses}) instead. The present article indirectly gives a positive answer to this question. Nevertheless, in any case it is Property \ref{decomp1DB} which is the key of both approaches.

\vspace{1cm}
\textbf{Acknowledgments}

The author would like to thank E. Guadagnini and Ph. Mathieu for many fruitful discussions.

\vfill\eject

\end{document}